\documentclass[12pt,english,a4paper]{article}

\usepackage[utf8]{inputenc}
\usepackage{indentfirst}
\usepackage{epsfig,calc}
\usepackage{amssymb,amsmath}
\usepackage{wrapfig}
\usepackage{hyperref}

\newcommand{\fdg}{.\!\!\degree} 
\makeatletter 

\renewcommand{\@makecaption}[2]{
	\vspace{\abovecaptionskip}%
	\sbox{\@tempboxa}{#1. #2}
	\ifdim \wd\@tempboxa >\hsize
	{\small{#1. #2}}\par
	\else
	\global\@minipagefalse
	\hbox to \hsize {\small{\hfil #1. #2\hfil}}%
	\fi
	\vspace{\belowcaptionskip}}

\newcommand{\degree}{^\circ}
\DeclareMathOperator{\diverg}{div}

\title{\raggedright{\normalsize{UDC 523(94+98)}}\\\centering{\textbf{Variation of global and local flows in the solar convection zone during activity cycles 24 and 25}}}

\author{A. V. Getling$^1$ and A. G. Kosovichev$^2$}
\date{}
\begin{document}
	\maketitle
	\begin{center}
		$^1$Skobeltsyn Institute of Nuclear Physics,\\ Lomonosov Moscow State University, Moscow, 119991, Russia;
		
		$^2$New Jersey Institute of Technology, Newark, NJ 07102, USA
	\end{center}
	\begin{abstract}
		Convection, differential rotation, and meridional circulation of solar plasma are studied based on helioseismic data covering the period from May 2010 to August 2024, significantly prolonged compared to that previously considered. Depth variation in the spatial spectrum of convective motions indicates a superposition of differently scaled flows. The giant-cell-scale component of the velocity field demonstrates a tendency to form meridionally elongated (possibly banana-shaped) structures. The integrated spectral power of the flows is anticorrelated with the solar-activity level in the near-surface layers and positively correlates with it in deeper layers. An extended 22-year cycle of zonal flows (``torsional oscillations'' of the Sun) and variations of the meridional flows are traced. A secondary meridional flow observed at the epoch of the maximum of Solar Cycle 24 to be directed equatorward in the subsurface layers is clearly manifest in Cycle 25.
	\end{abstract}
	
	\section{Introduction}\label{intr} 
	
	Flows in the Sun's convection zone play a crucial role in the origin of magnetic fields that command the entire chain of active processes occurring at various distances from the photosphere. This applies to both global streams---differential rotation and meridional circulation---and local convective flows. Because only photospheric layers are accessible for direct observation, helioseismic data are to be used to study the dynamics of the convection zone.
	
	Here, we present the results of our analyses of the horizontal velocity fields in the convection zone, based on Doppler measurements from the Helioseismic and Magnetic Imager (HMI) instrument of the orbital Solar Dynamics Observatory (SDO) and derived using time--distance helioseismological techniques \cite{Couvidat_etal_2012,Zhao_etal_2012} for the period from May 2010 to August 2024 with an 8-hour cadence for eight mean depth values of the layers\footnote{These data are expeditiously published at \url{http://jsoc.stanford.edu/data/timed/} \cite{Zhao_etal_2012}}. The data cover most of the 24th and virtually the complete ascending branch of the 25th solar cycle (previously, we used data for the entire period of SDO/HMI observations up to September 2020   \cite{Getling_etal_2021,Getling_Kosovichev_2022}). The source velocity-field maps are for the central part of the Sun's visible hemisphere, of size (in heliographic coordinates) $123\degree\times 123\degree$. The data refer to the following characteristic depths (the corresponding depth intervals in megameters are given in brackets): $d$ = 0.5 (0--1)~Mm, 2.0 (1--3)~Mm, 4.0 (3--5)~Mm, 6.0 (5--7)~Mm, 8.5 (7--10)~Mm, 11.5 (10--13)~Mm, 15.0 (13--17)~Mm, 19.0 (17--21)~Mm.
	
We describe the results of our study of subsurface convection in Section~\ref{conv} and then, global differential rotation and meridional flows in Sections~\ref{diffrot} and \ref{meridcirc}. Each section is preceded by general information on flow classification, including a historical background.
	
	\section{Convection}\label{conv} 
	
	\subsection{From granules to supergranules}
	
An important trait of solar convection is its multiscale structure. While granules with their characteristic size of about 1~Mm and lifetimes of a few minutes can easily be detected visually and have been known since the earliest telescopic observations of the Sun \cite{Herschel1800}, more subtle methods were required to identify larger structural elements of the convective velocity field.
	
\begin{figure}[t] 
		\centering{\includegraphics[width=0.8\textwidth,bb=0 0 340 340,clip]{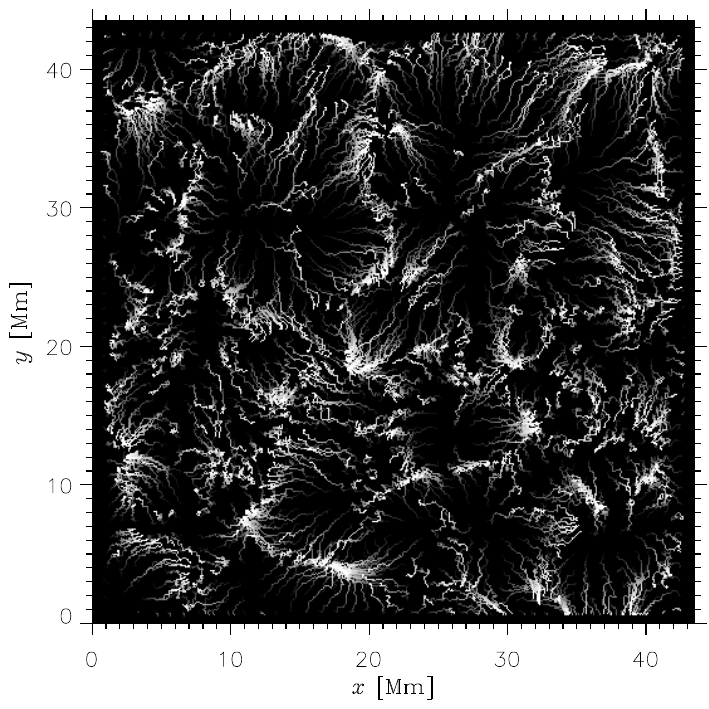}
		\caption{Cork trajectories obtained from a series of solar-photosphere images covering 1 h 45 min \cite{GetlingBuchnev2010}. Each trajectory is black in its initial section and merges with the background, its brightness gradually increasing to a maximum at the trajectory end. 
		The concentrations of the final segments of the trajectories mark the boundaries of supergranules and mesogranules.}\label{track}}
	\end{figure}

	Supergranules, whose sizes typically range between 20 and 30~Mm and whose lifetimes are of order 24 hours or longer, have been detected by Doppler measurements of horizontal velocities at some distance from the center of the solar disk (Hart and then Leighton et al. \cite{Hart_1954,Leighton_etal_1962}). Supergranulation can also be revealed by local correlation tracking of photospheric brightness-field elements (primarily granules). Fig.~\ref{track} was constructed using an improved tracking procedure in which ``corks'' (test particles) were selected based on the requirement of either maximum brightness-field contrast or maximum entropy of the brightness distribution \cite{GetlingBuchnev2010}. 
	
	In 1981, Doppler measurements of vertical velocities revealed convection structures on scales of 5--10~Mm, intermediate between granules and supergranules,---mesogranules (November et al. \cite{November_etal_1981}). Subsequently, doubts have been raised about the reality of mesogranulation, which were based on the lack of a corresponding peak in the convection power spectrum (whereas peaks are present on the granulation and supergranulation scales). Nevertheless, cork-trajectory maps like that shown 	in Fig.~\ref{track} delineate mesogranules quite clearly and leave no room for such doubts. The absence of the mesogranulation spectral peak can be explained by a wide spread of the mesogranule sizes.
	
	\subsection{Giant cells}\label{giant}
	
	In 1964, Bumba et al. \cite{Bumba_etal_1964} found evidence of ``giant structures'' in the distribution of the background magnetic fields in the photosphere. Next, Simon and Weiss \cite{Simon_Weiss_1968} predicted the existence of giant convection cells based on the location of the partial-ionization zones of hydrogen and 
	helium in the convection zone of the Sun; note that, as Glatzmaier and Gilman have shown \cite{Glatzmaier_Gilman_II_1981}
	(and should be expected from diverse studies of convection), giant cells extending through the entire thickness of the convection layer are the most easily excited mode of instability. However, detecting them in observations has proven challenging because their flow velocities are small compared to those in granulation and supergranulation cells. Therefore, giant cells have been referred to in the literature as hypothetical for three and a half decades. The first direct (Doppler) observations of giant cells 		became known only in 1998 (Beck {\cite{Beck_1998}). Subsequently, they have also been detected by tracking the motion of supergranules (Hathaway et al. \cite{Hathaway_etal_2013}).
				
The global scale of giant cells in the rotating Sun suggests that the Coriolis force substantially affects the flow in these cells. The degree of influence of rotation is determined by the Taylor number
	\begin{equation}
		\mathcal T=\frac{4\omega^2h^4}{\nu^2},
	\end{equation}\label{Taylor}
	where $\omega$ is the rotation rate of the fluid as a whole, $h$ is a characteristic scale of the flow region, and $\nu$ is the viscosity. If the Taylor number is sufficiently large, the conditions approach those of the Taylor--Proudman theorem, which prohibits three-dimensional flows  in a rotating volume in the limiting case of $\mathcal T\to\infty$, allowing only two-dimensional flows invariable in the direction of the rotational axis of the fluid as a whole. Then, if convection takes place in a spherical shell, meridionally aligned, curved  banana-shaped cells can be formed. Such a possibility has been repeatedly noted since the 1970s \cite{Busse_1970} (see also a review by Busse \cite{Busse_2002}). In particular, such cells were observed in laboratory experiments \cite{Busse_Carrigan_1974}. However, it is difficult to answer the question of how strong the effect of rotation on large-scale solar convection could be: it is necessary to know the effective turbulent viscosity, which cannot be estimated without knowing the scale and characteristic values of the velocity pulsations that determine it. We will see below that the spatial spectrum of convective flows indicates of the presence of cells stretched along meridians, so-called ``banana-shaped'' cells. 
	
	\subsection{Spatial spectrum of the convective velocity field}
	
	For some issues related to the multiscale structure of solar convection to be resolved, data for subphotospheric flows should be considered. In particular, the pattern of smaller-scale structures advected by larger-scale flows, observed in the photosphere \cite{Muller_etal_1992, Rieutord_etal_2001,GetlingBuchnev2010,Hathaway_2021}, in combination with hydrodynamic considerations \cite{Shcheritsa_etal_2018} suggests that the convective velocity field should appear as a superposition of differently scaled flows. On the other hand, the mixing-length theory used in the standard stellar evolution theory assumes that the flow at any given depth in the convection zone has a single   characteristic scale increasing with depth.
	It is therefore important to make a right, observationally justified choice between these possiibilities.
	
	Here, we present the spatial spectra of the convective horizontal-velocity field at different depths. For our spectral analysis, a scalar function, viz., the divergence of the horizontal velocity 
	\begin{equation}
		f(\theta,\varphi)=\diverg \mathbf V(\theta,\varphi),
	\end{equation} is more convenient than the vector of this velocity $\mathbf V(\theta,\varphi)$ (here, $\theta$ and $\varphi$ are the polar and the azimuthal angle in spherical coordinates). We represent $f(\theta,\varphi)$ as a spectral decomposition
	\begin{equation}
		f(\theta,\varphi) = \sum_{\ell=0}^{\ell_{\max}}\sum_{m=-\ell}^{\ell}A_{\ell m}Y_{\ell}^{m}(\theta,\varphi)
		\label{series}
	\end{equation}
	in spherical harmonics of degree $\ell$ and azimuthal number $m$
	(here, $\ell_{\max}$ is a suitably chosen upper spectral boundary). The wavelength of the harmonic  $Y_{\ell}^{m}(\theta,\varphi)$ on a sphere of radius $r$ is determined by the Jeans formula \cite{Jeans_1923}
	\begin{equation}
		\lambda=\frac{2\pi r}{\sqrt{\ell(\ell+1)}}.
		\label{Jeanseq}
	\end{equation}
	 	We use it to estimate the characteristic scales of convection structures, assuming $r$ to be equal to the solar radius $R_\odot$ (this is justified by the small thickness of the convection-zone shell under consideration compared to $R_\odot$). For a given $\lambda$, these characteristic scales can differ depending on the configuration of the structures, being of order $\lambda$; for example, in the case of roll convection, the roll width is $\lambda /2$, whereas, for a system of hexagonal convective cells, the size of each cell is $(2/\sqrt 3)\lambda$ in one direction and $2\lambda$ in the other, perpendicular to the first direction.
	
	We will be interested in the power spectrum of the flow
	\begin{equation}\label{power}
		p_{\ell m}=|A_{\ell m}|^2.
	\end{equation}
According to Parseval's theorem, the integrated power of the flow described by the function $f(\theta,\varphi)$ can be written as
	\begin{equation}
		p_\mathrm{tot} \equiv \int\limits_\Omega f^2 \mathrm d\Omega =\sum_{\ell=0}^\infty \sum_{m=-\ell}^\ell |A_{\ell m}|^2
	\end{equation}
	(here, $\mathrm d\Omega$ is an elementary and $\Omega$ is the full solid angle).
	
	An accurate spectral representation of granulation requires a very large $\ell_{\max}$, at which the result is highly noisy. We smooth the original velocity field with a 17.5-Mm window, thus eliminating the granulation. We take $\ell_{\max}=201$ as the upper limit of the spectrum.
	
	\begin{figure}
		\centering{
			\includegraphics[width=0.4\textwidth,bb=20 0 780 650, clip]{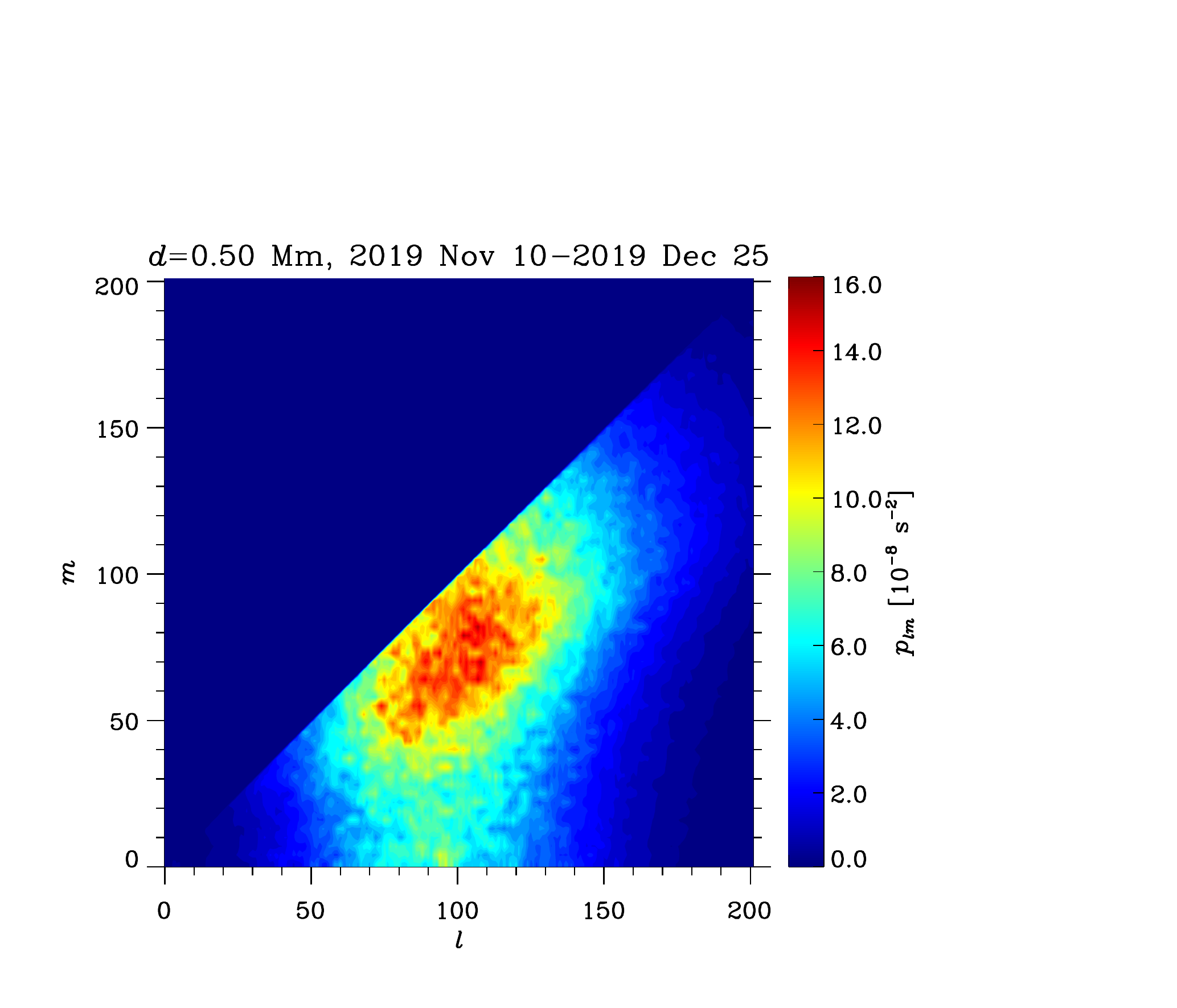}
			\includegraphics[width=0.4\textwidth,bb=20 0 780 650, clip]{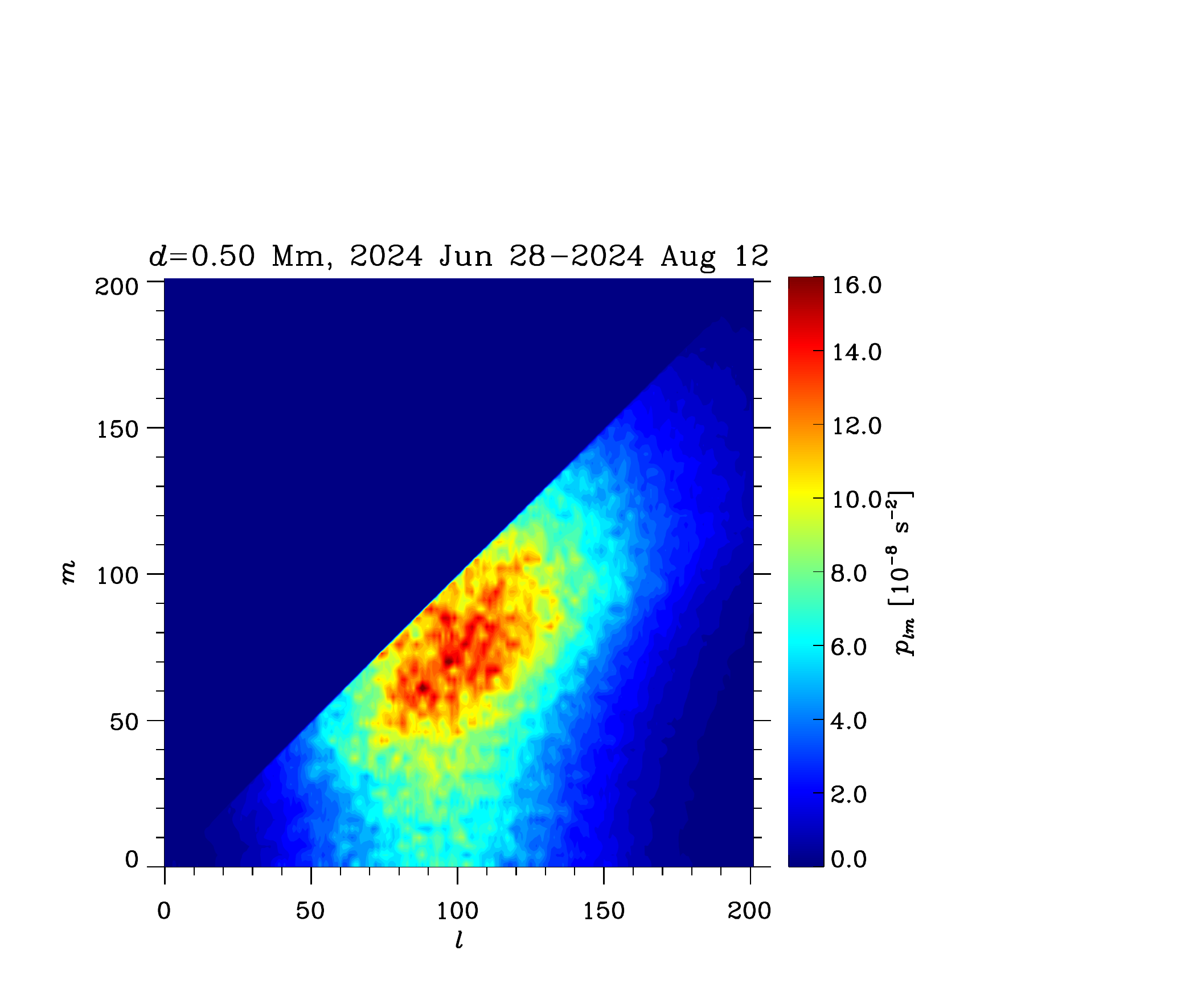}\\
			\includegraphics[width=0.4\textwidth,bb=20 0 780 650, clip]{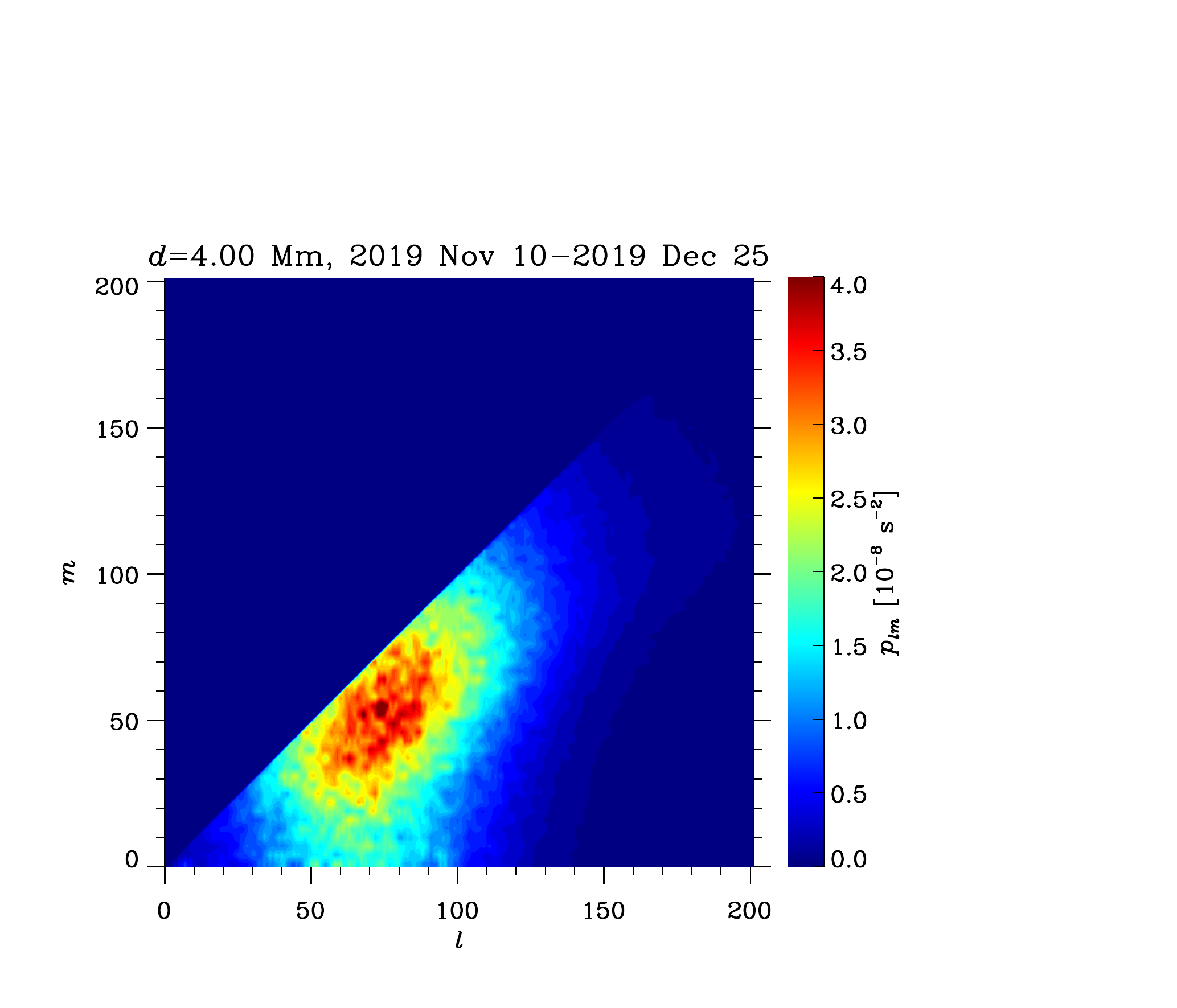}
			\includegraphics[width=0.4\textwidth,bb=20 0 780 650, clip]{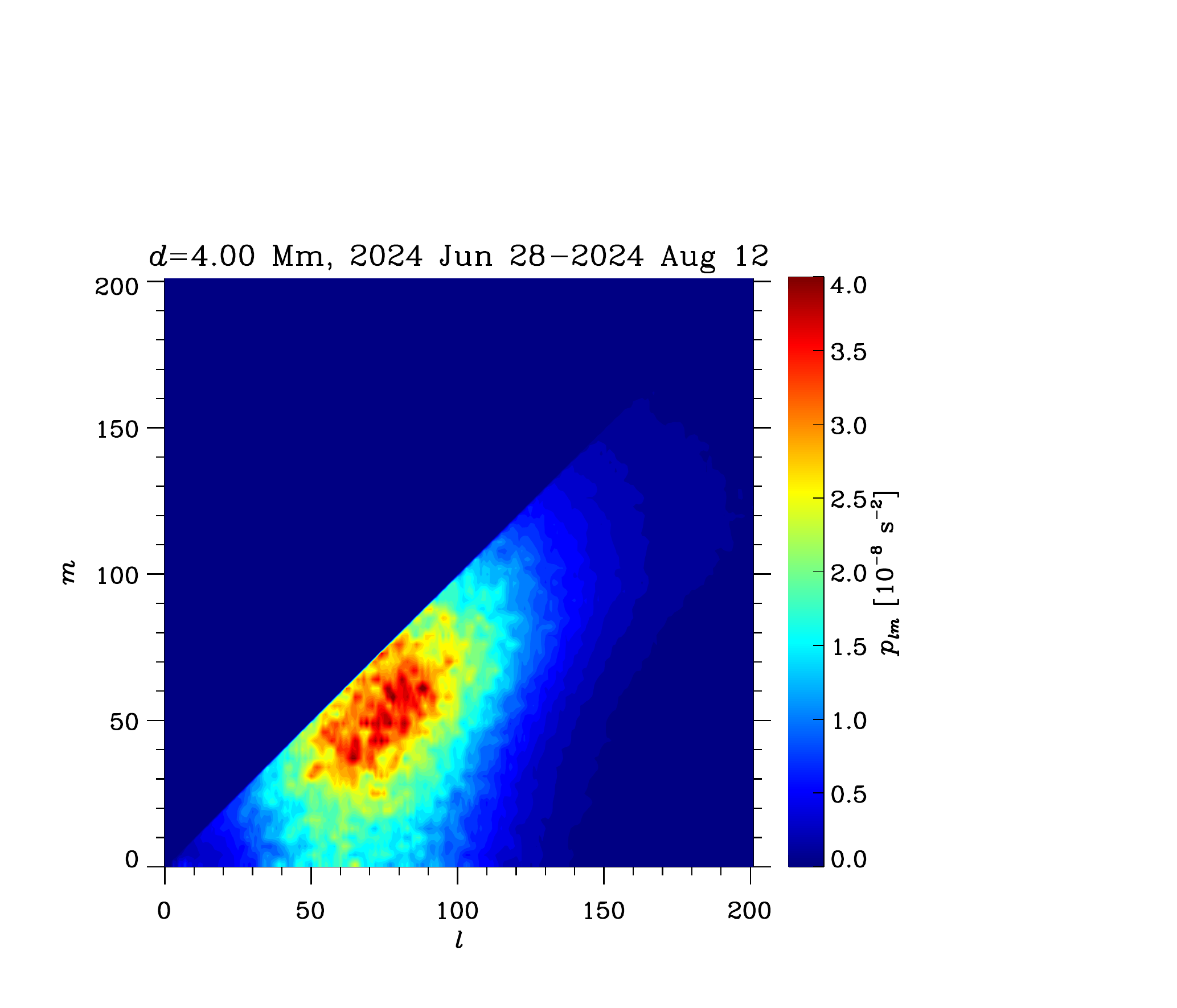}\\
			\includegraphics[width=0.4\textwidth,bb=20 0 780 650, clip]{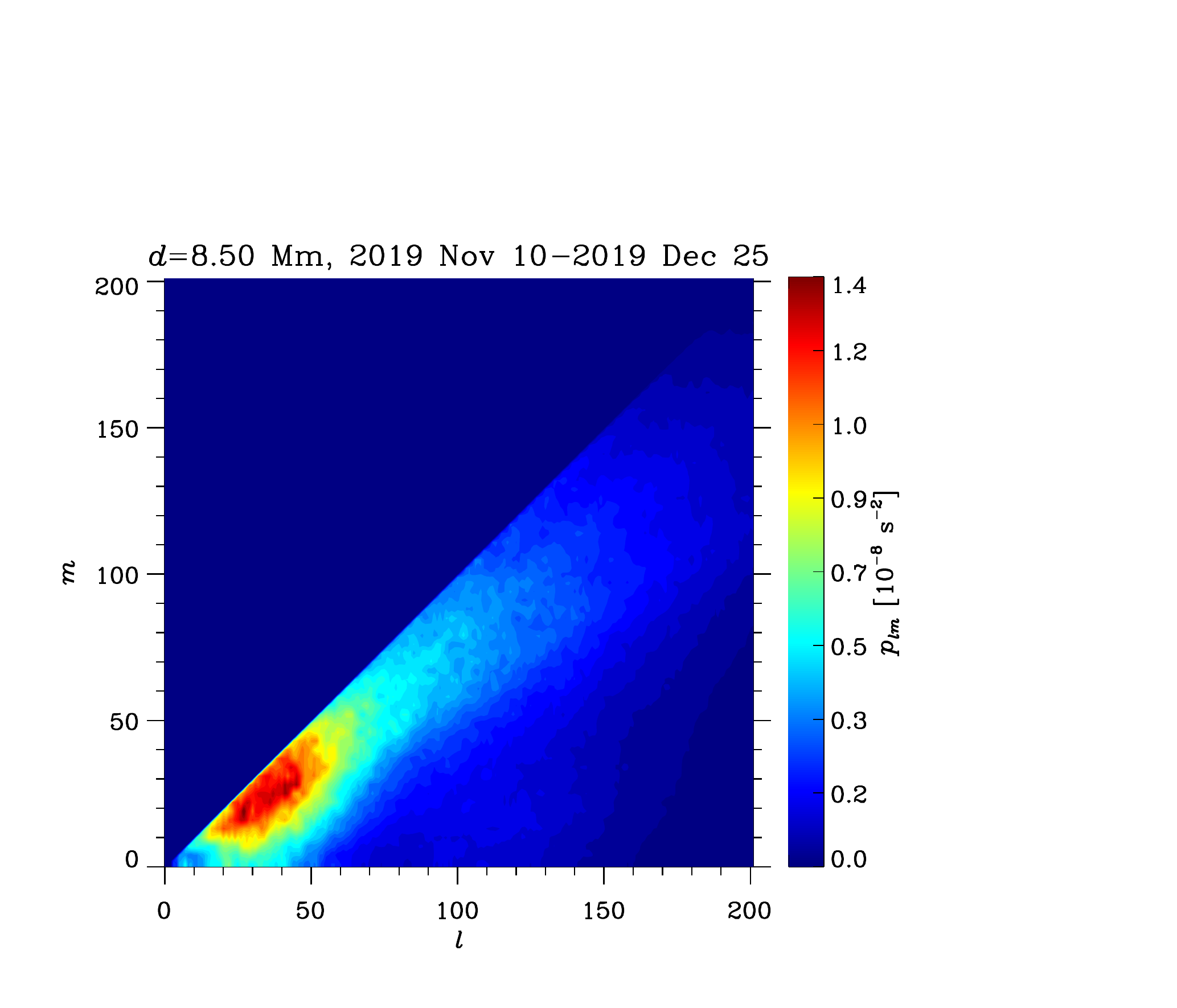}
			\includegraphics[width=0.4\textwidth,bb=20 0 780 650, clip]{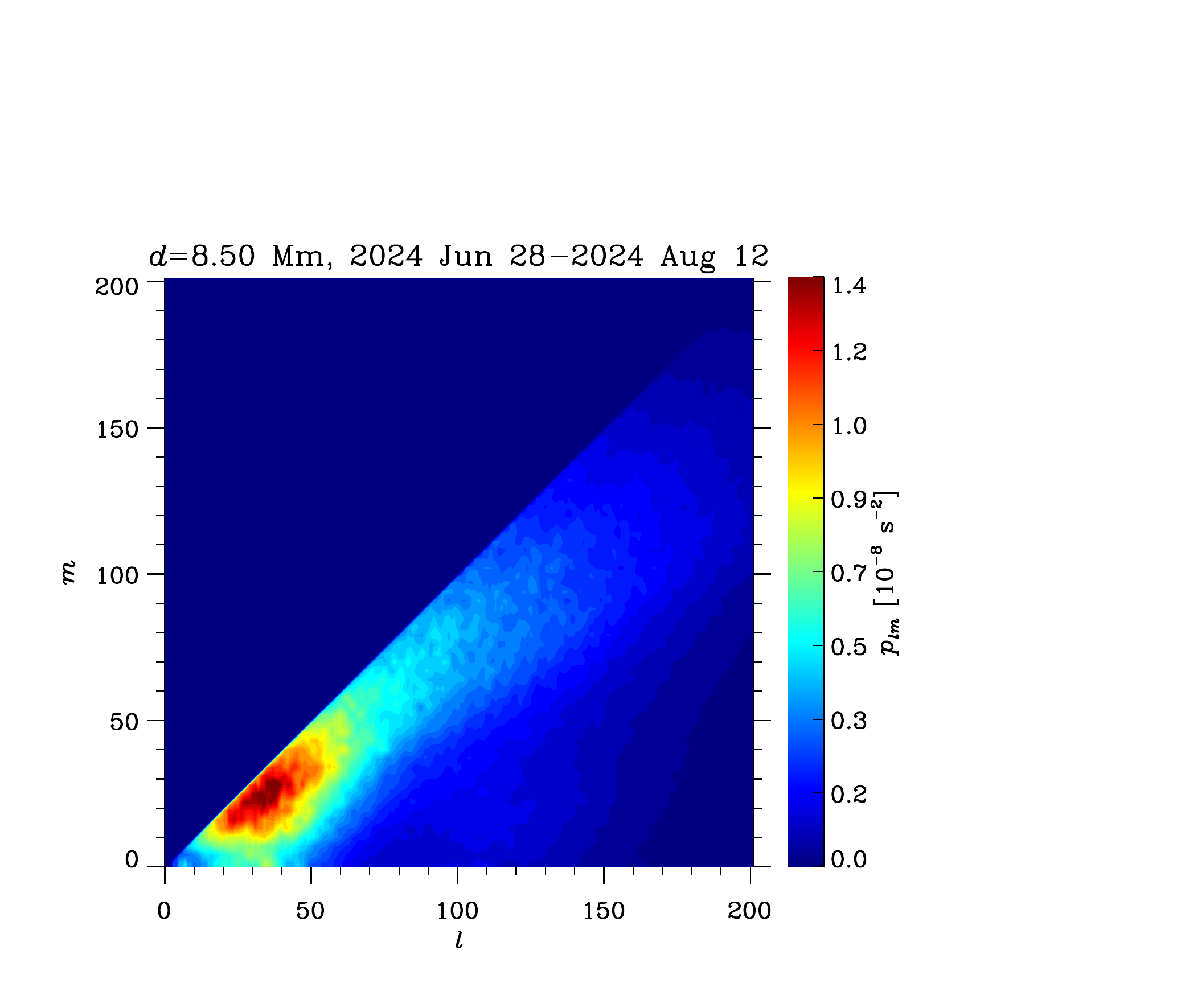}\\
			\includegraphics[width=0.4\textwidth,bb=20 0 780 650, clip]{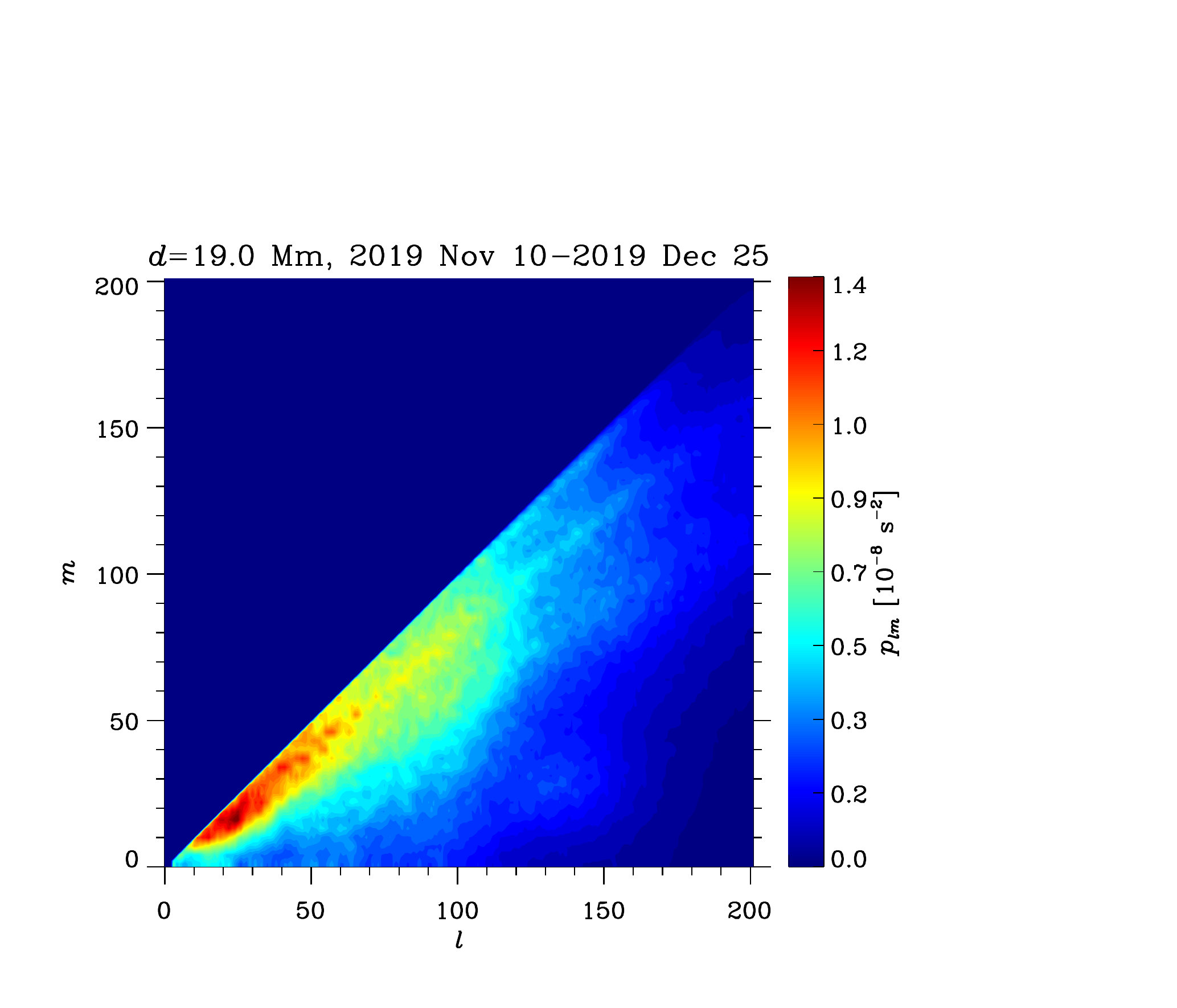}
			\includegraphics[width=0.4\textwidth,bb=20 0 780 650, clip]{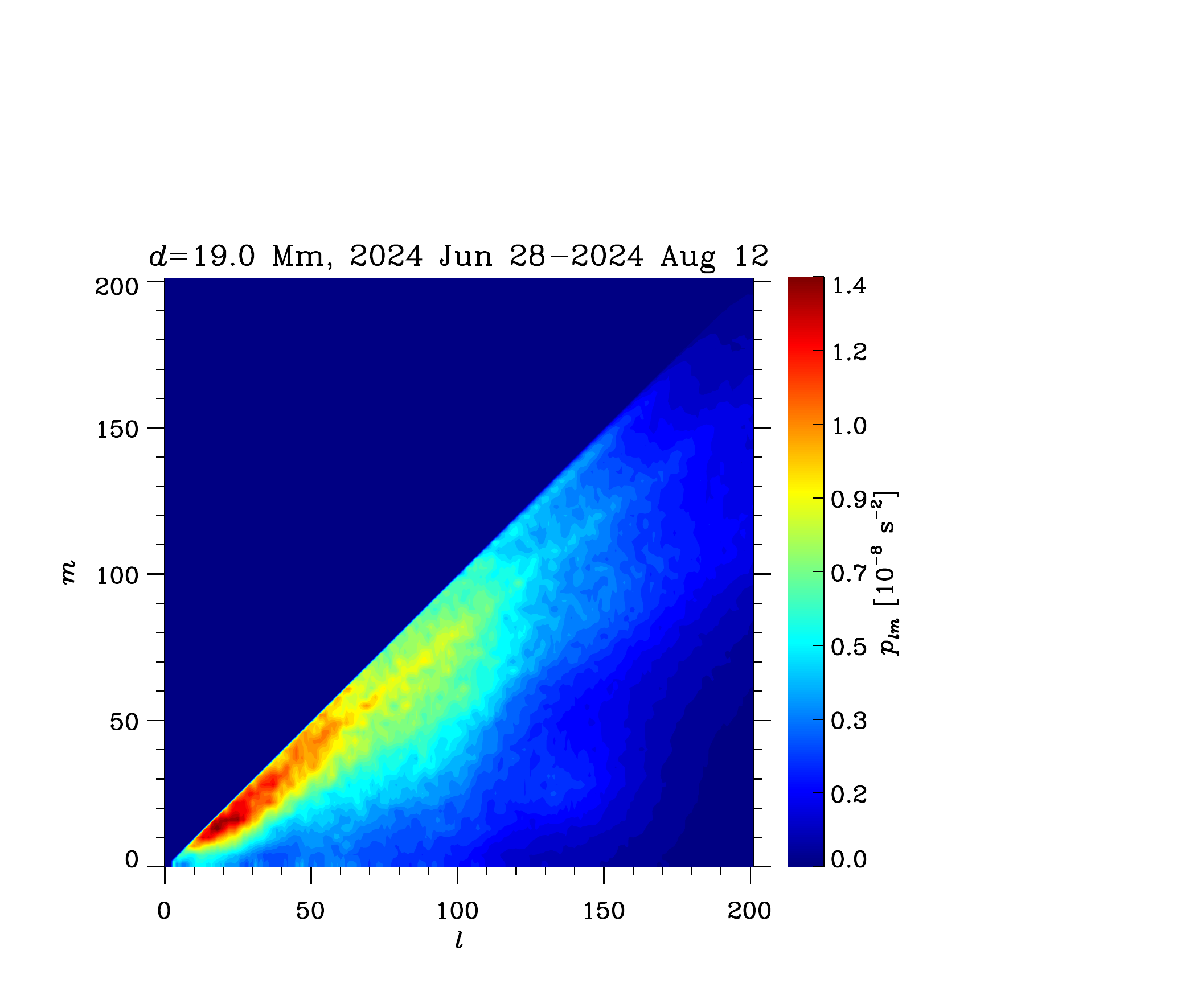}
			\caption{Depth variation of the 45-day-averaged power spectrum $p_{lm}$ of the horizontal-velocity divergence field (the selected periods, as well as depths $d$, are indicated at the top of each panel). Left column: low-activity period November 10--December 25, 2019; right column: high-activity period June 28--August 12, 2024.}\label{spectra}}
	\end{figure}
	
	
	Since the data that we use do not cover the whole spherical surface, we cut a $120\degree$-wide longitudinal sector  out of each flow map and complement it with the same data shifted in longitude by $120\degree$ and $240\degree$.
	The resultant spectra contain only harmonics with non-zero $m$ multiples of 3. We interpolate these spectra to all missing $m$ values and smooth them with a two-point window for better visual perceptibility. 

To reduce the effects of the Gibbs phenomenon, we smooth the transition in latitude $\phi$ to the ``empty'' polar caps
 $-90\degree \leqslant \phi < -61\fdg 5$ and $61\fdg 5 < \phi \leqslant 90\degree$  (where data are missing due to the choice of the dimensions of the area under study, $123\degree\times123\degree$).  
As follows from an experiment with model fields, ``empty'' polar caps only slightly narrow the bandwidth of $\ell$ values; smoothing the cap boundaries also has a minimal effect on the spectrum \cite{Getling_Kosovichev_2022}.

A more detailed description of the analysis procedure and its accuracy checks is given in our earlier paper \cite{Getling_Kosovichev_2022}.

Typical flow spectra for different depths are shown in Fig.~\ref{spectra}. Notably, the main spectral peak shifts to small $\ell$'s and narrows with the increase of depth. In the upper layers, it corresponds to $\ell \sim 70$--$130$, $\lambda \sim 30$--$60$~Mm, which are  supergranulation scales. At the bottom layer boundary, the most energetic harmonics correspond to $\ell \sim 20$, so that the center of the wavelength band (very broad absolutely but more narrow relatively) lies near 300~Mm---this is a giant-cell scale. 

At $d=19$~Mm, a typical spectrum is peaked near $\ell=22$. It is remarkable that the corresponding value of the $m$-averaged power, $\langle p_{22,m}\rangle_m$, varies little over the depth range considered: near the surface, it is approx\-imately 5/8 of its peak value for $d = 19$~Mm. This means that large-scale structures are also present in the near-surface velocity field, although they are hardly distinguishable against the background of more energetic, smaller-scale flows. Consequently, large-scale flows extend through a large depth range, and the velocity field is a superposition of differently scaled flows.

Another important feature of the convection spectrum is that the main spectral peak approaches the $m=\ell$ line as the depth is increased. Since the $m=\ell$ harmonics are sectorial, i.e., at a given longitude, they increase monotonically in each hemisphere from the pole to the equator as $\sin^m\theta$, this means that large-scale convection tends to form meridionally elongated, ``banana-shaped'' cells.

\subsection{Variation of the integrated power of convection over the activity cycle}

As we have already noted \cite{Getling_Kosovichev_2022}, the integrated convection power $p_\mathrm{tot}$
experiences variation over the 11-year solar activity cycle. The revealed regularities manifest even more clearly on the time interval extended until the maximum of Cycle 25. To remove semi-annual fluctuations, apparently due to changes in the inclination of the Sun's rotational axis to the line of sight, the time variation of the integrated power is Fourier filtered with an ideal low-pass filter
$$H(\nu)=\left\{
\begin{aligned}
	&1,\quad \nu \leqslant \nu_\mathrm H,\\
	&0, \quad \nu > \nu_\mathrm H,
\end{aligned}
\right.$$
and with a Butterworth low-pass filter
$$H(\nu)=\frac{1}{1+(\nu/\nu_\mathrm H)^{2n}};$$
here $\nu_\mathrm H$ is the cutoff frequency assumed to be 0.05~$\mu$Hz, and $n=4$ is chosen.

The variations found in this way and presented in Fig. \ref{timevar} reveal a distinct anticorrelation of the flow power with the  solar-activity level in the upper layers and an equally distinct positive correlation at greater depths. The depth variation of the correlation coefficient is shown in Fig.~\ref{correl}. Such a behavior of the convection power can be interpreted as a result of the subsurface magnetic field suppressing convection in the upper 
layers and redistributing the flow energy into deeper layers, where it can weaken turbulent fluctuations and thereby reduce eddy viscosity.

\begin{figure}
	\centering{
		\includegraphics[width=0.425\textwidth,bb=0 0 566 283, clip]{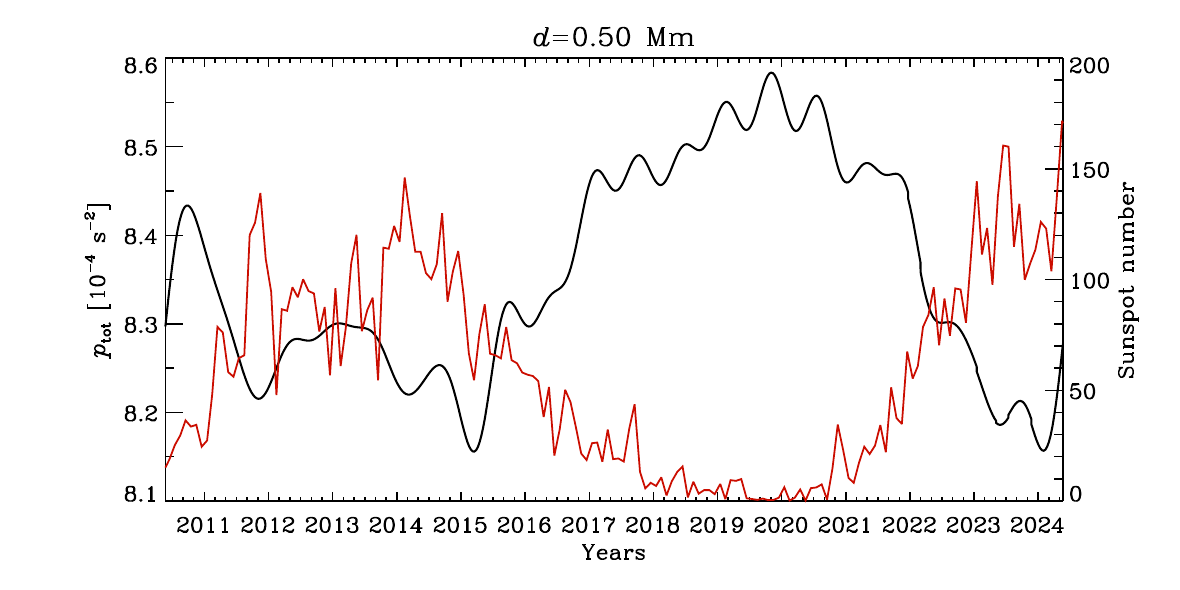}
		\includegraphics[width=0.425\textwidth,bb=0 0 566 283, clip]{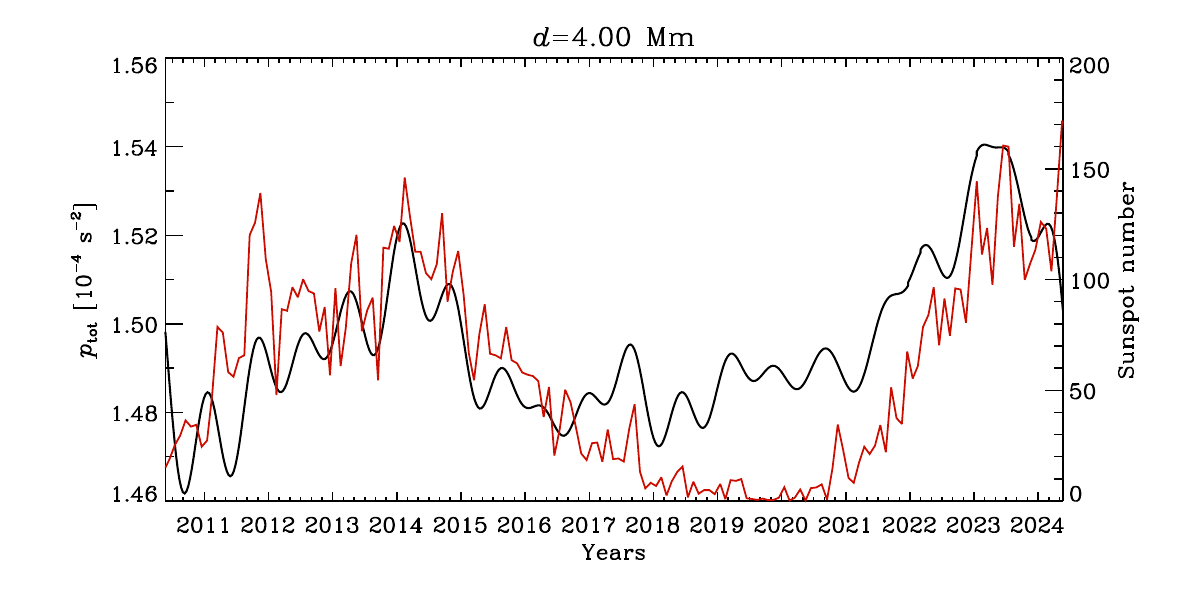}\\
		\includegraphics[width=0.425\textwidth,bb=0 0 566 283, clip]{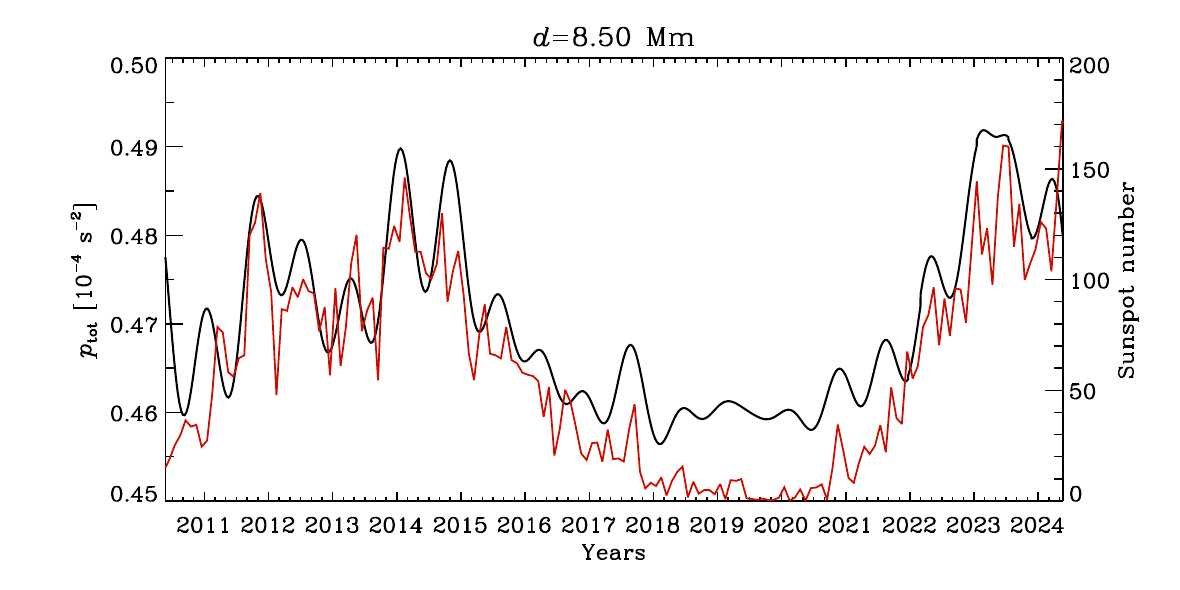}
		\includegraphics[width=0.425\textwidth,bb=0 0 566 283, clip]{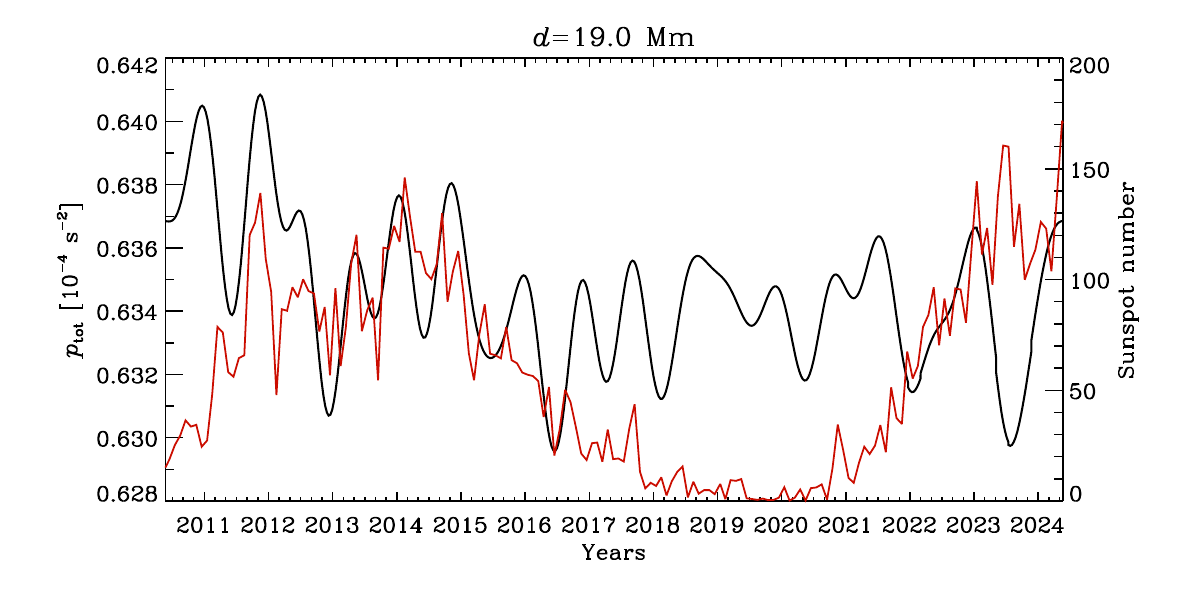}
		\caption{Time variation of the total spectral power of the horizontal-velocity divergence field at different depths (black curves). The semi-annual variations due to the changes in the inclination of the Sun's rotational axis to the line of sight are eliminated by spectral filtering. The red curves show the variation of the monthly averaged  sunspot number. Depth values are indicated at the top of each panel. }\label{timevar}}
\end{figure}

\begin{figure}[t] 
	\centering{\includegraphics[width=0.6\textwidth]{	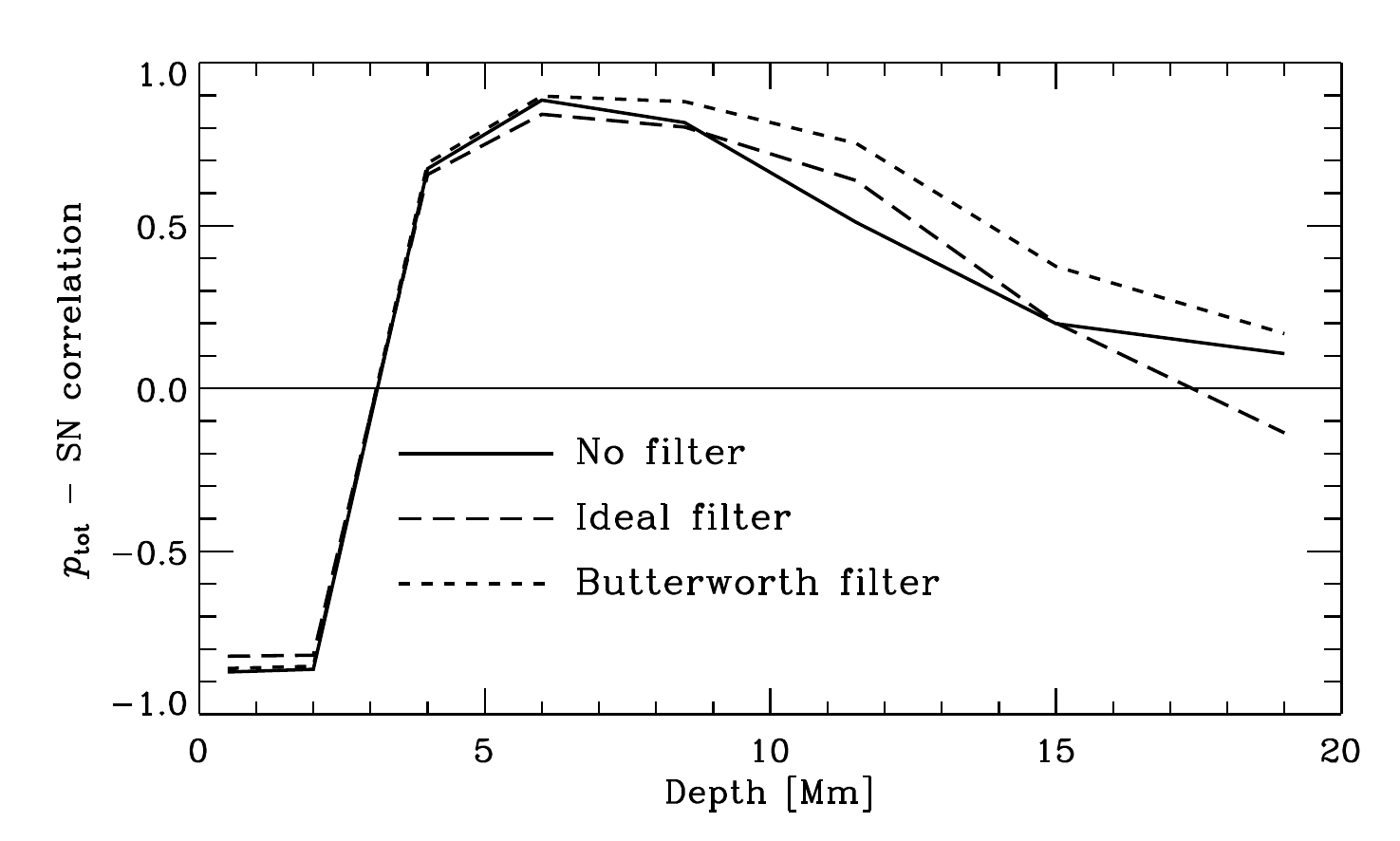}\caption{Correlation of the integrated spectral power of convection with the solar-activity level. }\label{correl}}
\end{figure}

\begin{figure}[t] \centering{\includegraphics[width=0.8\textwidth]{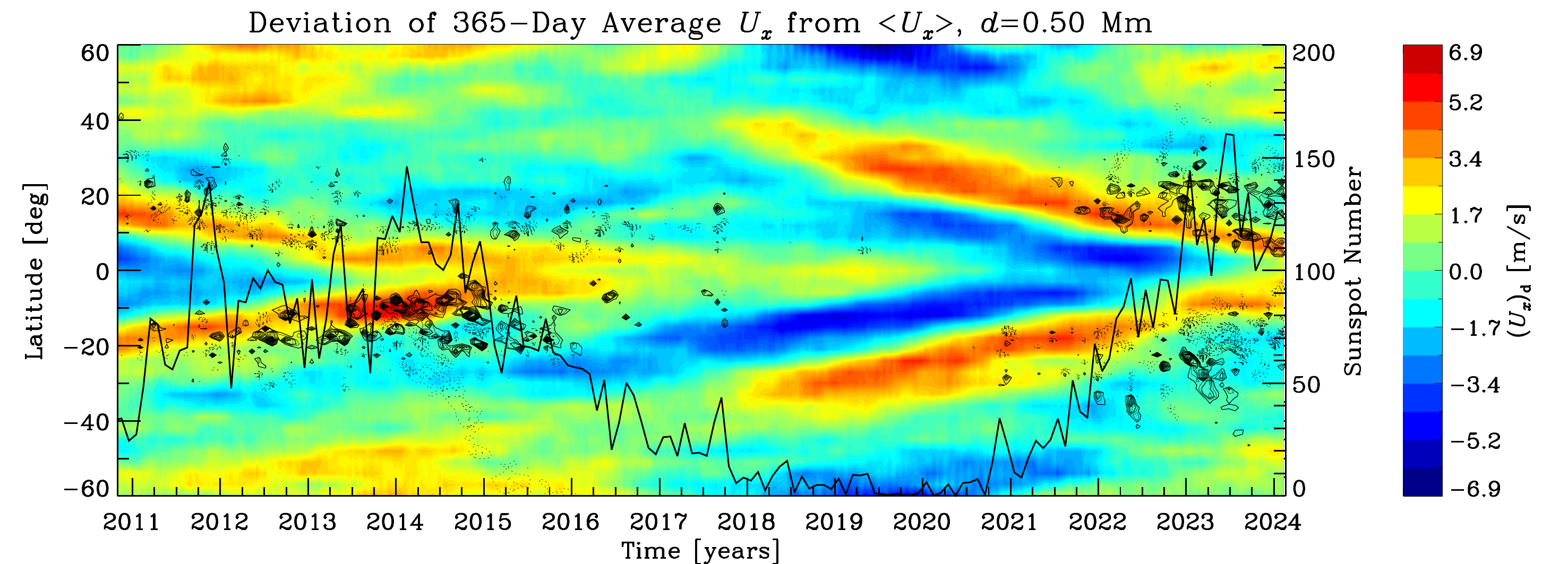}\\
		\includegraphics[width=0.8\textwidth]{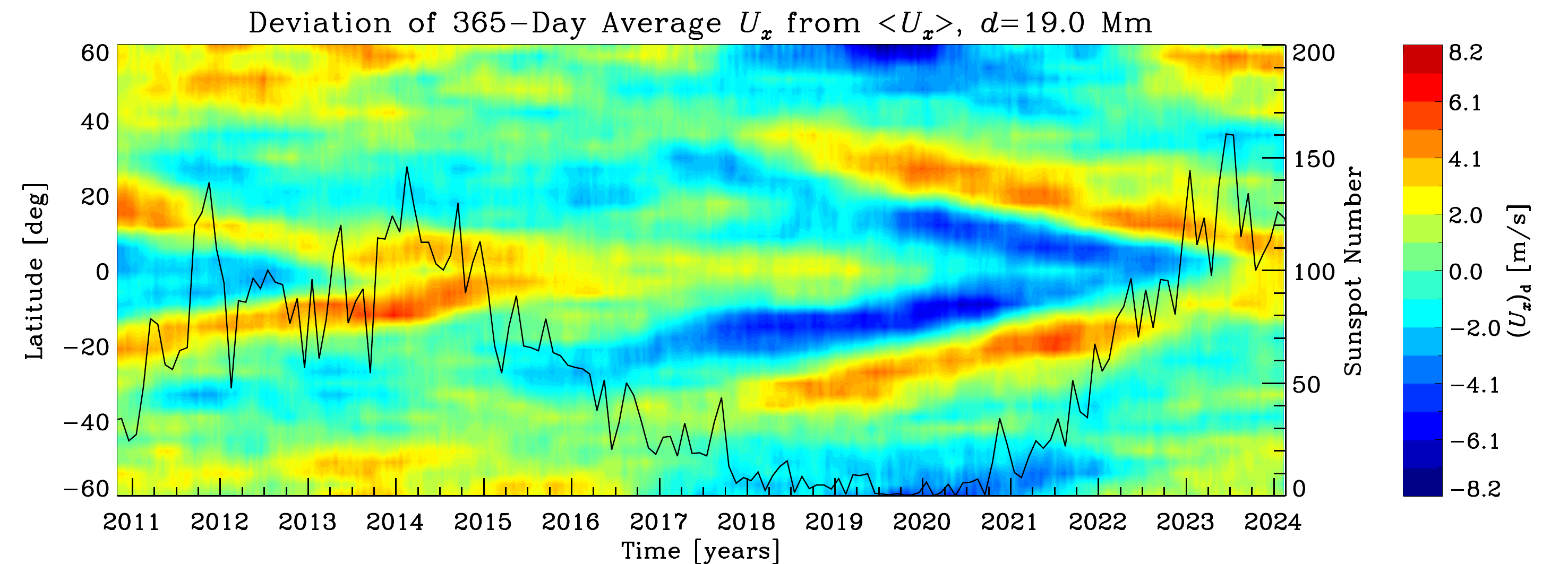}
		\caption{Differential rotation and torsional oscillations of the Sun according to helioseismic data. We present time--latitude diagrams for the deviations of the azimuthal component of the solar-plasma velocity at depths of $d=0.5$ and 19~Mm from its mean values for 2010--2024, obtained by applying moving averaging with a window of 365 days to remove the effects of annual changes in the inclination of the solar rotational axis to the line of sight. The black curve shows the changes in the monthly averaged sunspot number. In the top panel, the contours of the monthly averaged vertical component of the magnetic field are plotted in black.}\label{torsion}}
\end{figure}

\begin{figure}[t] \centering{\includegraphics[width=0.8\textwidth]{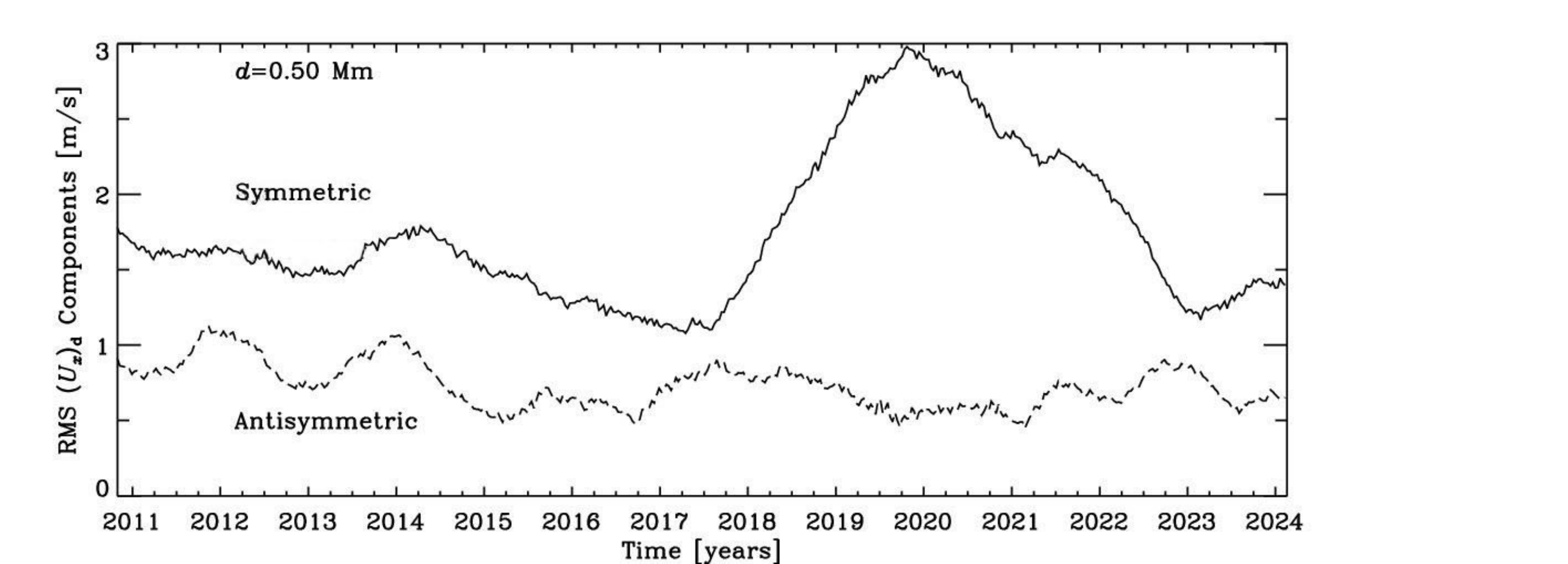}
		\includegraphics[width=0.8\textwidth]{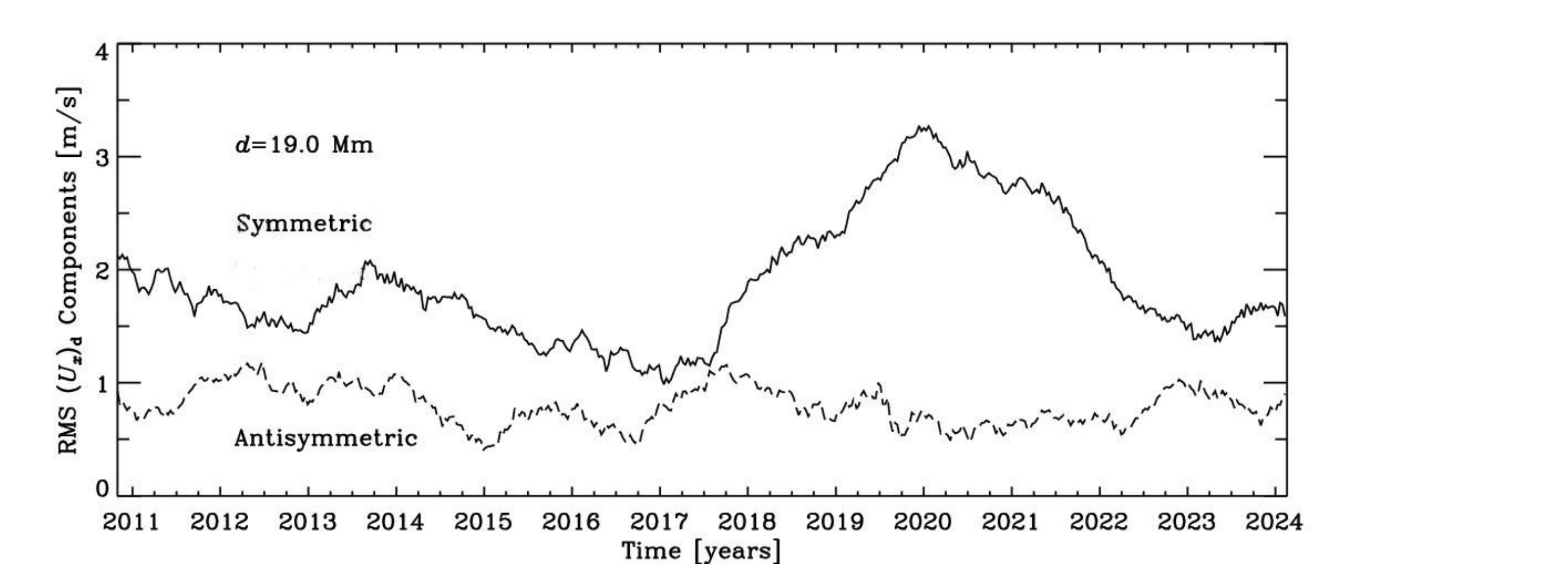}
		\caption{Variations (over the activity cycle) in the rms symmetric and antisymmetric components of the deviation of the zonal velocity $U_x$ (averaged with a 365-day moving window) from its mean value for 2010--2024 at depths of $d=0.5$ and 19~Mm.}\label{simmasimm}}
\end{figure}

\section{Differential rotation}\label{diffrot}

The differential rotation of the Sun (the decrease of the rotation rate from the equator to the poles) was discovered by Scheiner in 1630 from the motion of spots, and the first quantitative determnation of the rotation law, also based on observations of spots, belongs to Carrington \cite{Carrington_1863}. This law was substantially refined by Newton and Nunn \cite{Newton_Nunn_1951} using the motion of spots and Snodgrass \cite{Snodgrass_1983} by tracking magnetic fields. The Sun's rotation was further investigated by Doppler measurements \cite{Howard_etal_1983,Snodgrass_Ulrich_1990}, magnetic tracking---in particular, \cite{Meunier_1999,Zhao_etal_2004}---and using helioseismological methods, e.~g., \cite{Thompson_etal_1996}.

The most successful theoretical description of differential rotation considers the interaction between convection and global rotation: the Coriolis force affects turbulent convection, and convection, in turn, transfers the angular momentum giving rise to variations of rotation rate with depth and latitude. The idea of such interaction was first put forward by Lebedinsky \cite{Lebedinsky_1941} (see, in particular, a review by Kitchatinov \cite{Kitchatinov_2005}). 

\begin{figure}
	\centering{\includegraphics[width=0.8\textwidth]{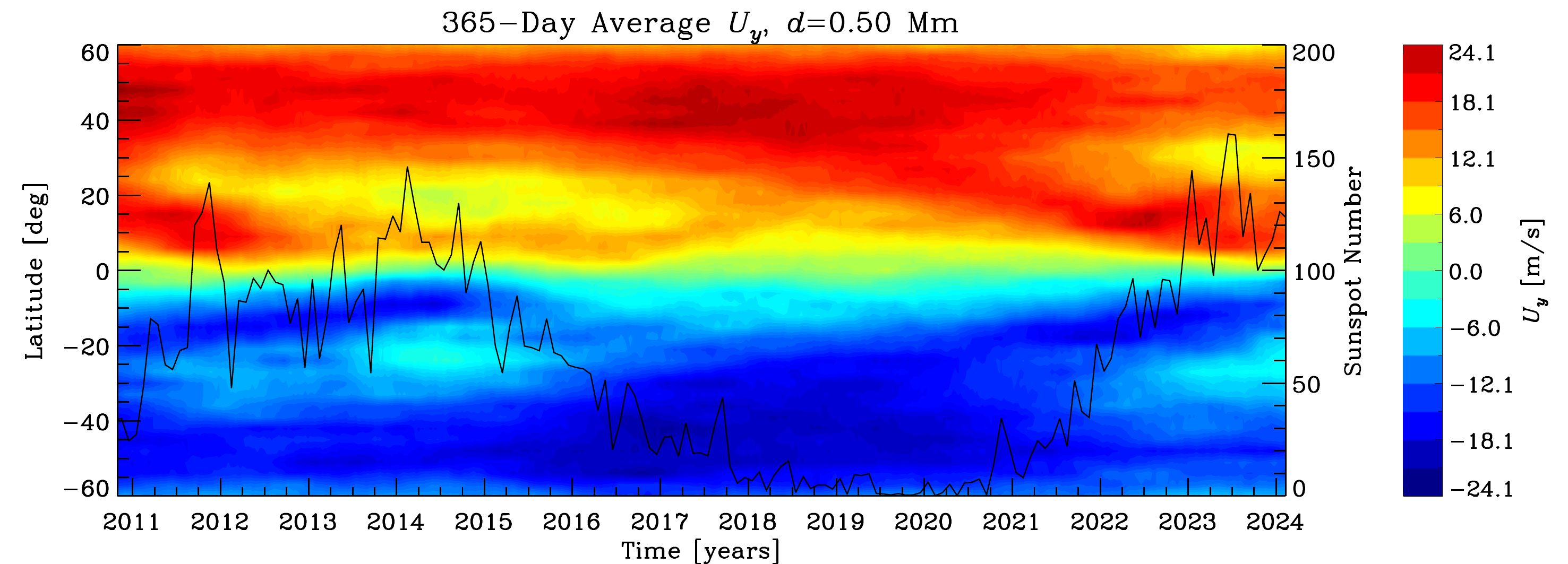}\\
		\includegraphics[width=0.8\textwidth] 
		{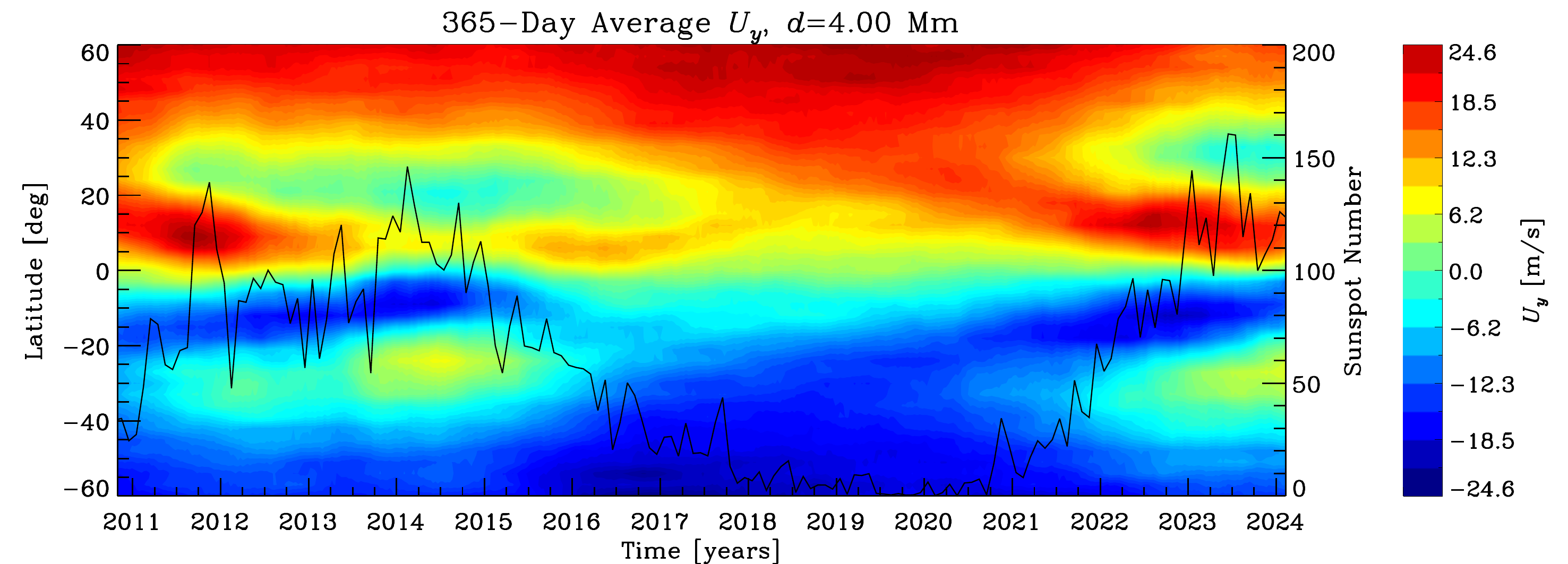}\\
		\includegraphics[width=0.8\textwidth] 
		{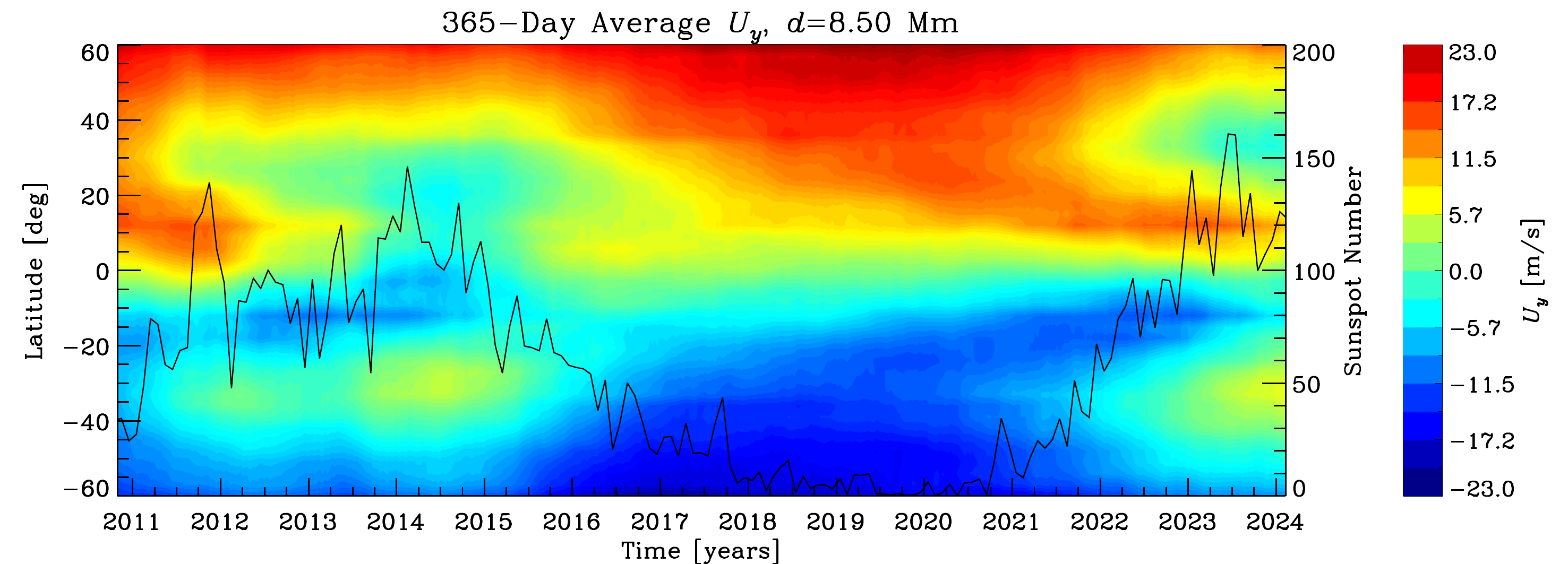}\\
		\includegraphics[width=0.8\textwidth] 
		{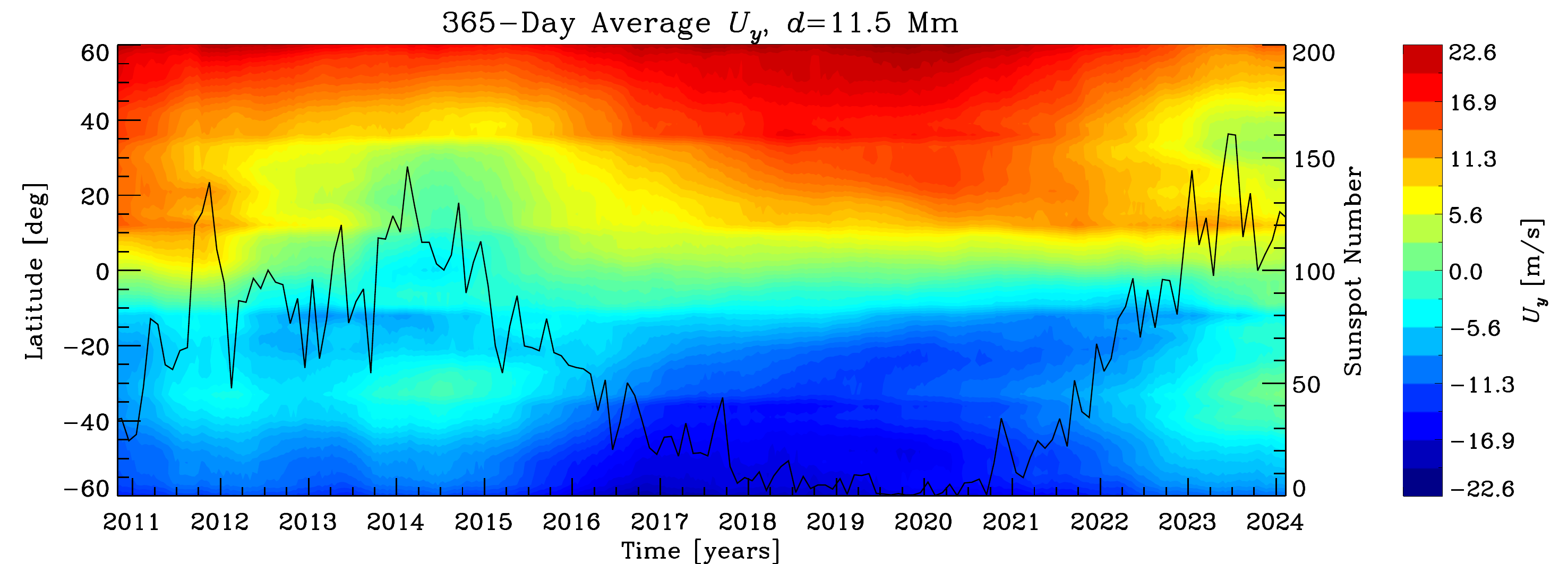}\\
		\includegraphics[width=0.8\textwidth] 
		{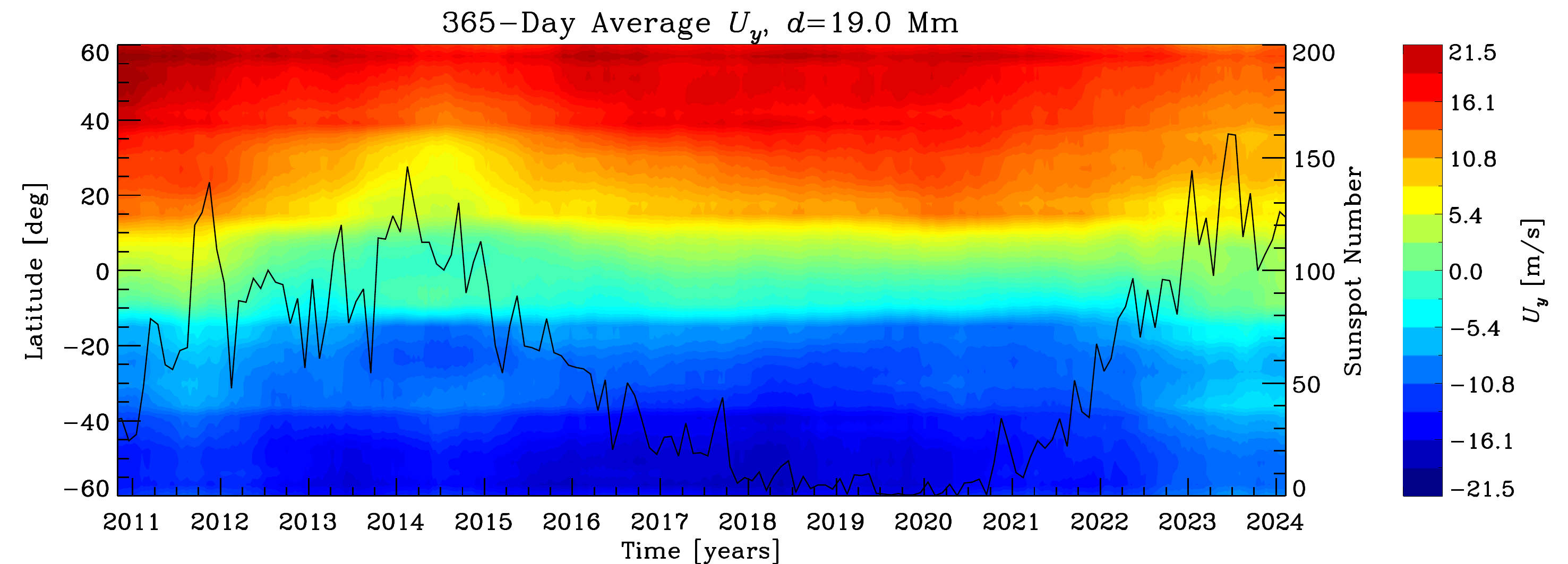}\caption{Time--latitude diagram for the meridional-flow velocity at different depths obtained by applying moving time averaging. The red curve shows changes in the monthly mean sunspot number.}\label{merid}}
\end{figure}
\begin{figure}
	\centering{\includegraphics[width=0.8\textwidth]{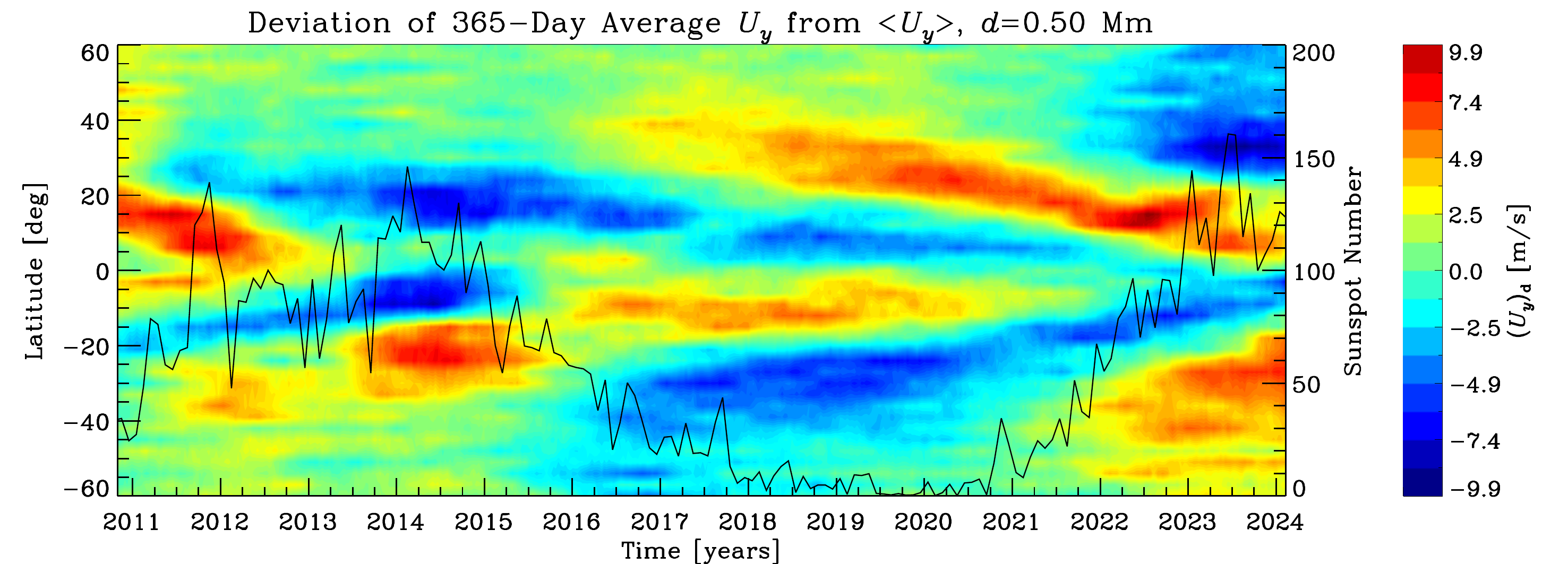}\\
		\includegraphics[width=0.8\textwidth] 
		{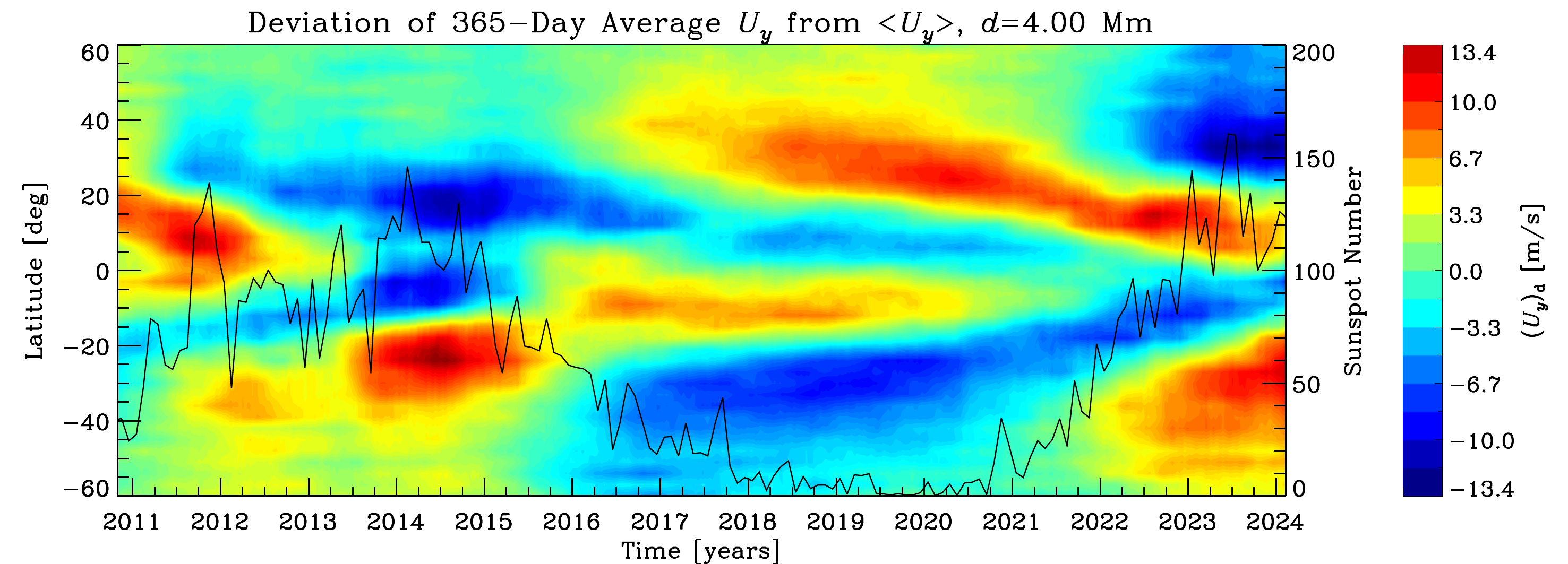}\\
		\includegraphics[width=0.8\textwidth] 
		{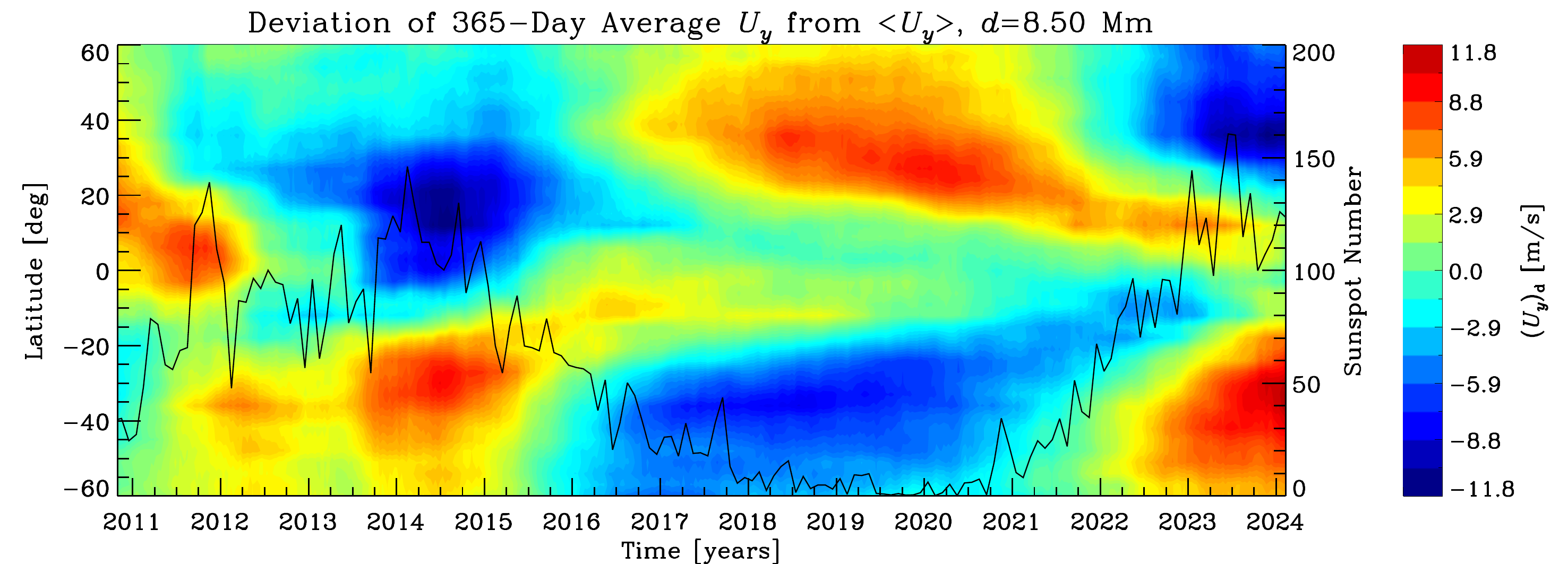}\\
		\includegraphics[width=0.8\textwidth] 
		{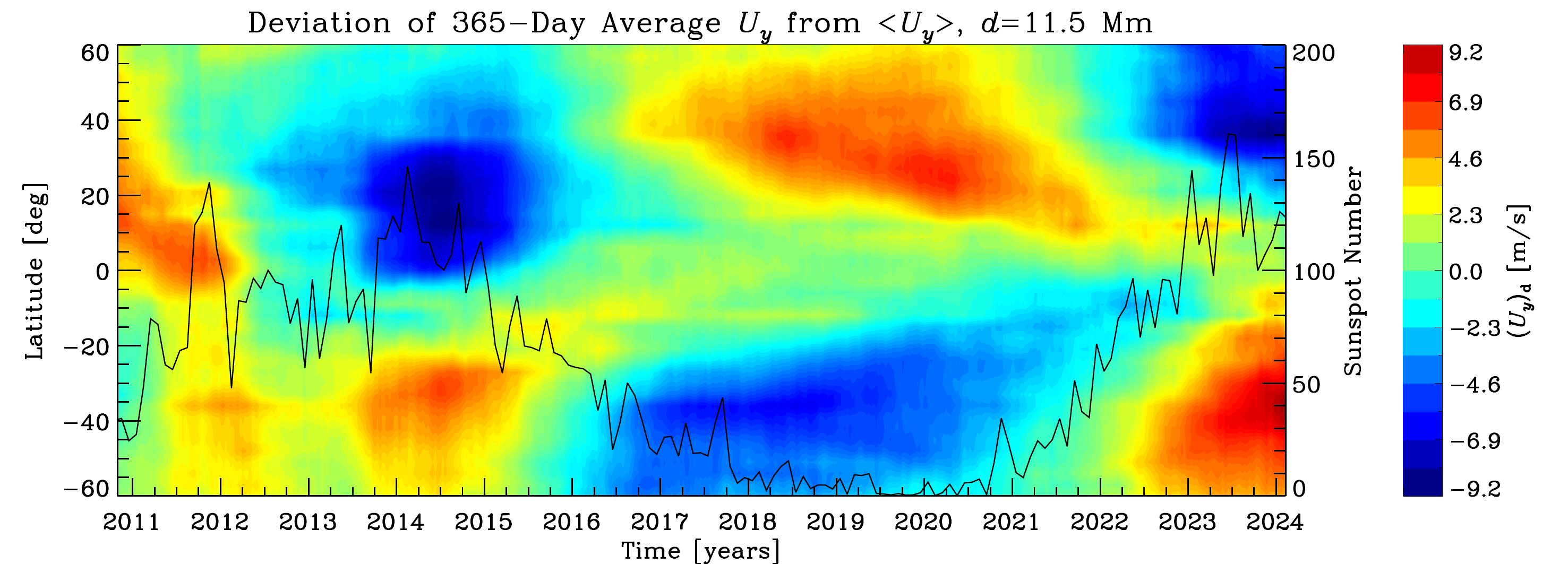}\\
		\includegraphics[width=0.8\textwidth] 
		{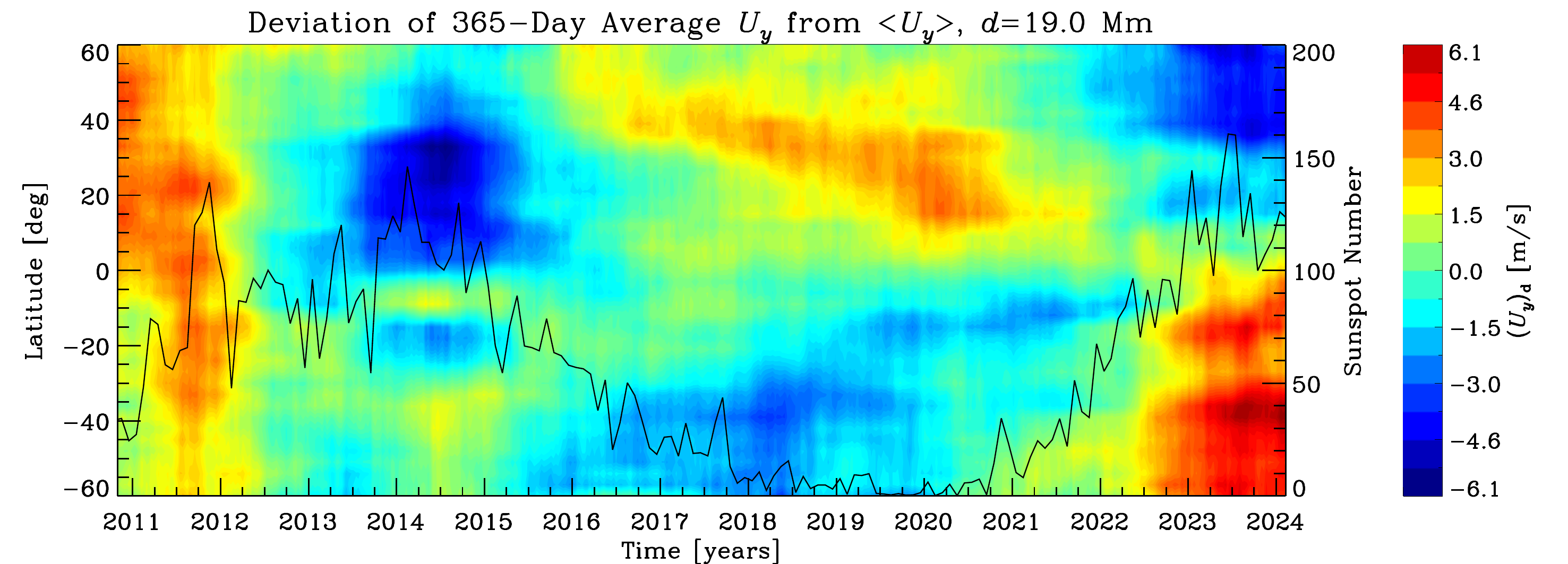}\caption{Time--latitude plots for the deviation of the annual moving-window average of the meridional-flow velocity from its mean value for 2010--2024 at various depths indicated in the figure. The black curve shows the changes in the monthly mean	 sunspot number. }\label{meriddev}}
\end{figure}

\begin{figure}[!t]
	\centering{
		\includegraphics[width=0.8\textwidth]{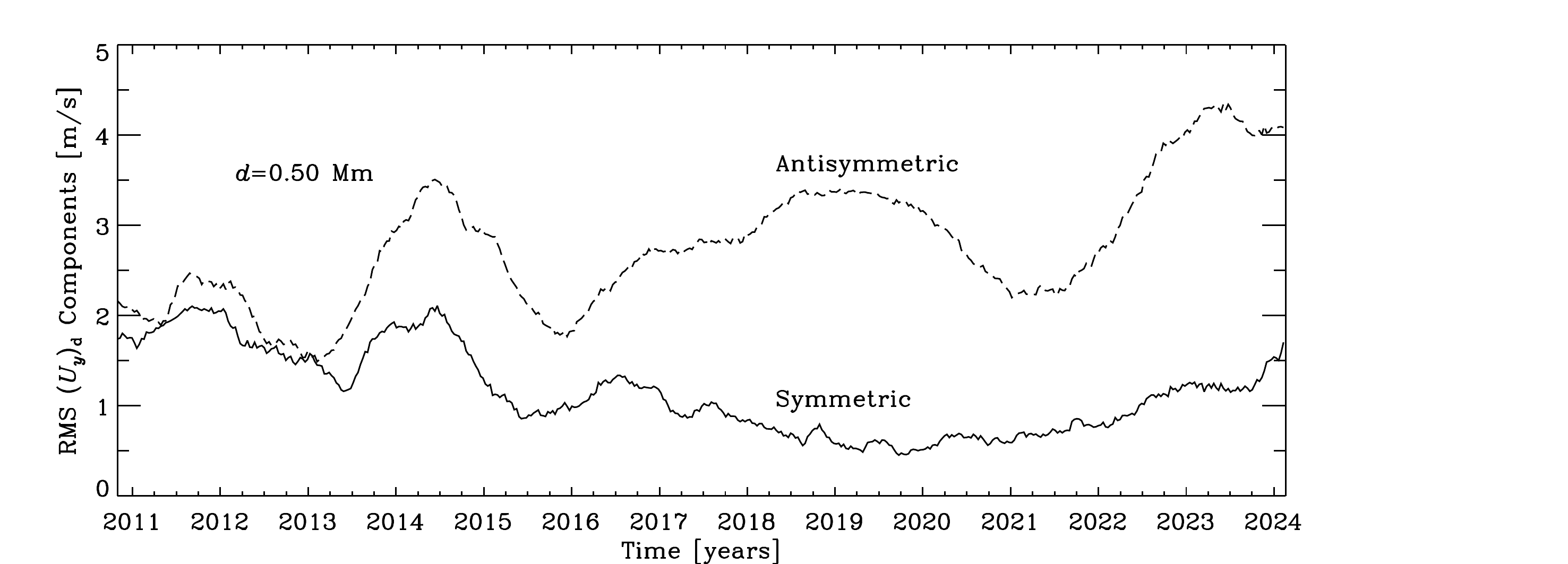}\\
		\includegraphics[width=0.8\textwidth]{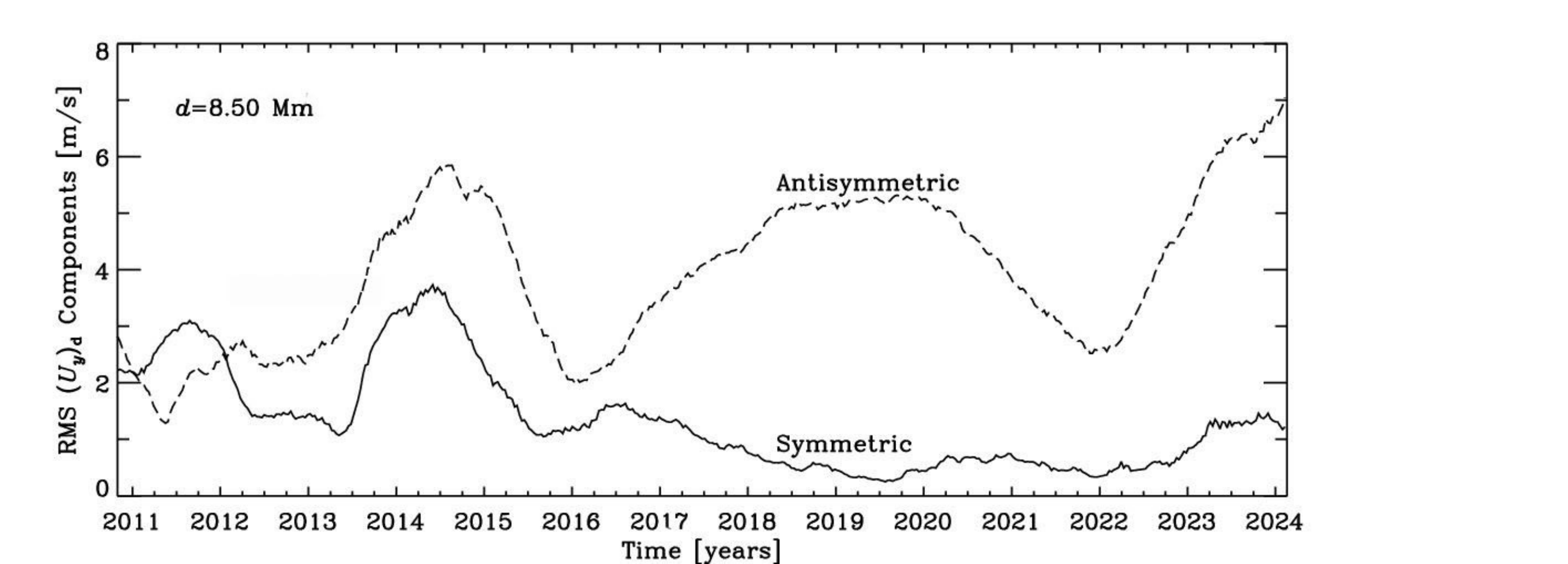}\\
		\includegraphics[width=0.8\textwidth]{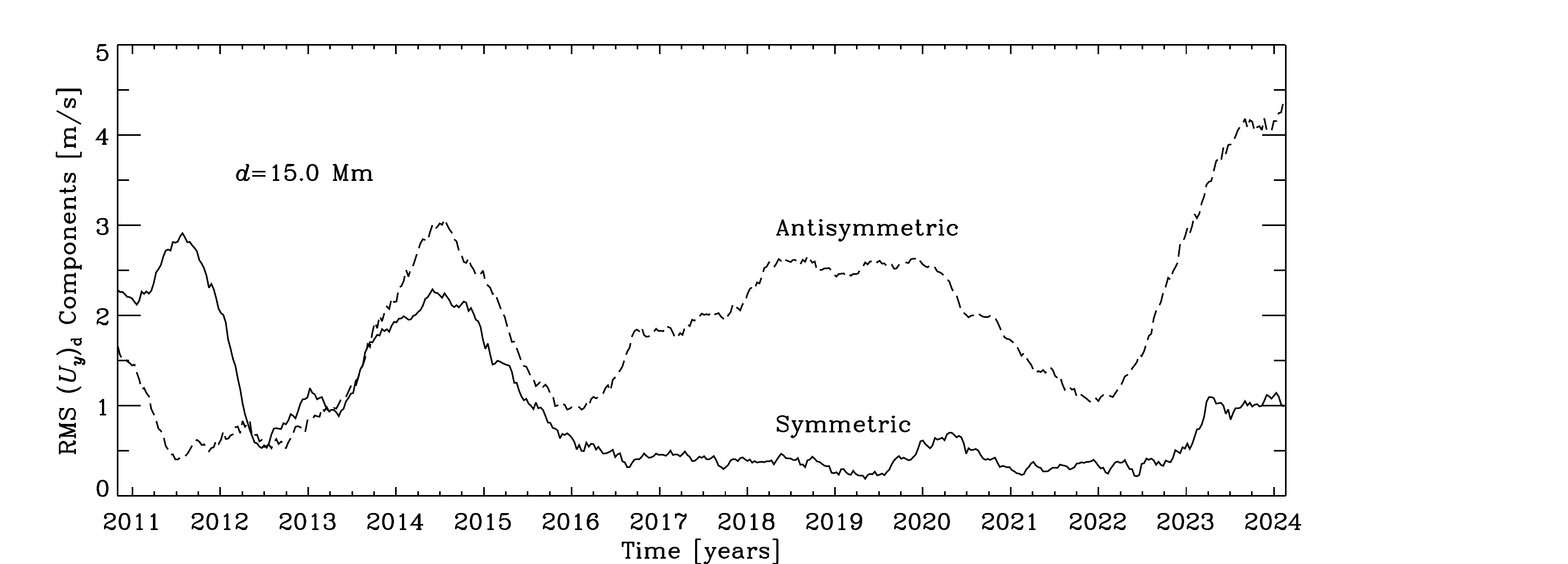}\\
		\includegraphics[width=0.8\textwidth]{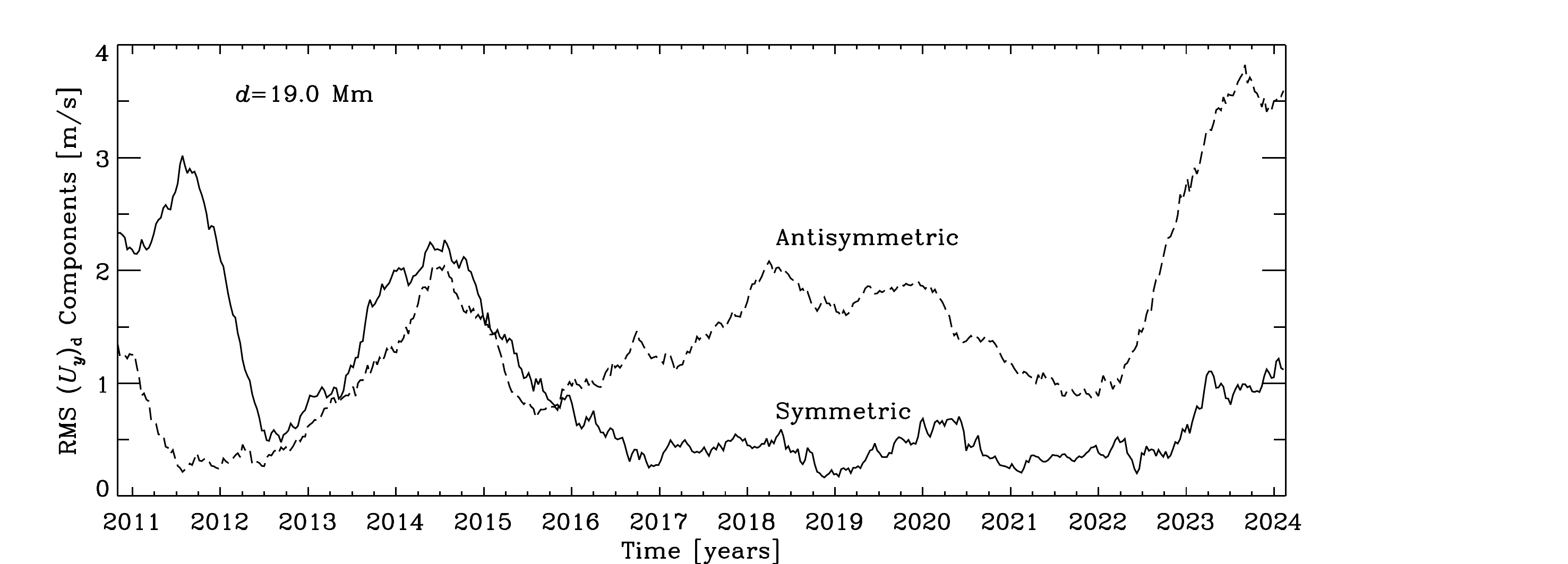}
		\caption{Variation of the RMS symmetric and antisymmetric components of the deviation of the annual-moving-window averaged meridional velocity $U_y$ from its mean value for 2010--2024 at depths $d=0.5$, 8.5, 15, and 19~Mm.}
		\label{meriddev_simasim}}
\end{figure}

Differential rotation varies with time. Howard and Labonte \cite{Howard_Labonte_1980}, using Doppler velocity measurements, discovered traveling torsional waves---altern\-ating latitudinal zones of accelerated and decelerated rotation. These zones originate at high latitudes and drift to the equator, where they disappear. The period of such variation is nearly 22 years and is known as the `extended' solar cycle \cite{Wilson_1988}. Torsional oscillations were further studied by Labonte and Howard \cite{Labonte_Howard_1982}, Snodgrass \cite{Snodgrass_1991}, etc.

Our study of differential rotation is based on the helioseismologically determined horizontal velocity of matter. The values of $v_x$ in the Carrington frame of reference for any time and any level $d$ were averaged within each of 41 latitudinal zone of width 1$\fdg92$, which were chosen in the latitude range from $-60\degree$ to $+60\degree$ with a step of $3\degree$. The obtained values were averaged over longitude and then their moving time average $U_x$ was calculated with a window of 365 days and a step of 10 days.  

Fig.~\ref{torsion} shows the time--latitude diagrams for the deviation of the rotational velocities at depths $d=0.5$ and 19~Mm from their mean values. They, along with the diagrams for the intermediate layers, differ little from each other. Since the diagrams cover a time interval shorter than 22 years, the entire extended cycle of torsional oscillations cannot be traced therein, as it was done by Kosovichev and Pipin using the data of global helioseismology \cite{Kosovichev_2019}. Nevertheless, the character of rotation-rate variation can be seen quite clearly. Bands of accelerated rotation, wavelike traveling in both hemispheres over latitude toward the equator, originate at high latitudes during the period of low activity and disappear, giving way to the decelerated rotation, two 11-year cycles later, shortly after the activity minimum. The origin of the zonal bands of slowed rotation cannot be traced so clearly in these diagrams because of the insufficient duration of the observation period; their disappearance at the equator occurs near the activity maximum.

	The azimuthal velocity field is not quite symmetric about the equator, which is not possible to find using the techniques of global helioseismology. Fig.~\ref{simmasimm} shows the time variation of the root-mean-square deviations of the symmetric and antisymmetric components of the annually averaged azimuthal velocity from their mean values over the entire period considered. These dependences are similar at all depths. It can be seen that the symmetric component reaches its largest values in the periods of low solar activity, while the antisymmetric component changes insignificantly.

\section{Meridional circulation}\label{meridcirc}

Kippenhahn has demonstrated \cite{Kippenhahn_1963} that differential rotation inevitably entails meridional circulation. Both these velocity fields are closely interacting; they cannot exist without each other and should be considered two sides of a single process.

As shown by Doppler measurements and tracking the motion of magnetic elements, the plasma on the solar surface flows from the equator to the poles with velocities of order 10--20~m/s. The velocity of this meridional flow grows with latitude from zero at the equator; it reaches a maximum near latitudes $\pm 39\degree$ and slowly decreases toward the poles. The tracking of sunspots, granules, and supergranules has also been used to investigate meridional flows. The solar plasma transported by the meridional circulation to the polar regions should sink into the deep convective zone, where a return flow to the equator and an upward flow in the equatorial zone should occur. Helioseismic data have significantly complemented surface measurements (see, e.~g., \cite{Beck_1998,Zhao_etal_2004,Zhao_2013,Gizon_etal_2020}). However, the detailed spatial structure of meridional flows is a matter of debate, especially for large depths, due to the low velocity of the flows.

The study of differential rotation and meridional circulation is especially important because modern theories of the solar dynamo include differential rotation as a fundamentally important ingredient of the dynamo mechanism, and satisfactory agreement of theoretical models with observations can be achieved only by taking into account the meridional circulation.

Our study of meridional flows uses the values of velocity, $v_y$, from the considered set of helioseismic data, and the data analysis is quite similar to that used to study the differential rotation. The time--latitude diagrams for the longitudinally averaged and time-averaged meridional velocity, $U_y$, with a window of 365 days are shown in Fig.~\ref{merid}.

	It can be seen that one peculiarity in the velocity distribution, which was observed at the epoch of the Cycle-24 maximum and noted by us earlier \cite{Getling_etal_2021}, repeats itself almost identically near the Cycle-25 maximum. Namely, in addition to the regular poleward flows in both hemispheres, there are regions of reduced velocities in the $d=0.5$-Mm layer near latitudes of $\pm 20\degree$, mainly in 2013--2015 and 2023--2024. At $d=4$--8.5~Mm, where this feature is even more pronounced, we observe a reversal of the meridional flows. On the contrary, near a latitude of $\phi=-10\degree$, the flow from the equator to the pole is accelerated in 2014, with a synchronous and symmetric velocity increase in the two hemispheres observed in 2023, and in the 24th activity cycle, the northern hemisphere outpaces the southern hemisphere, showing acceleration in 2011--2012. This pattern blurs and weakens with depth, almost completely disappearing by $d=19$~Mm. Thus, during periods of high solar activity, a higher harmonic is superposed on the usual pattern of meridional circulation, which in the near-surface layers corresponds to the additional flow converging toward the latitudinal zones of the most active sunspot formation, which was previously found in \cite{Haber_2002,Zhao_2004}.

Fig.~\ref{meriddev} presents the time--latitude diagrams for the deviation of the velocity, $U_y$, from its mean value for the entire period of 2010--2024. The diagrams show a pattern of differently directed velocities in the near-surface layers similar to the torsional-oscillation diagram (but, unlike it, odd relative to the equator); this similarity is especially noticeable in the upper layers. Thus, the changes in the meridional flows, similarly to the variations of differential rotation, follow an extended 22-year cycle.

The variation in the symmetric and antisymmetric components of the deviation of $U_y$ from its mean value are shown in Fig.~\ref{meriddev_simasim}. In the upper layers, the peaks of the main (antisymmetric) component occur in the years of both activity maxima and minima. In deeper layers, the peak of 2011--2012  disappears, and a peak of the symmetric component grows in the same time interval. No such phenomenon is observed in 2019--2020 minimum.

The coherence of the extended cycles of differential rotation and meridional circulation can be explained on the basis of the nonlinear model of dynamo and large-scale flows in the convection zone proposed by Pipin and Kosovichev \cite{Pipin_Kosovichev_2019}  using the techniques of mean-field magnetohydrodynamics \cite{Pipin_2008}. In this model:
\begin{itemize}
	\item the average convection velocity is calculated using the mixing-length approximation;
	\item heat transport is calculated with the rotation ($\Lambda$-effect) and magnetic-field effects taken into account;
	\item angular-momentum balance is described in terms of the $\Lambda$-effect theory using the anelastic approximation;
	\item mixing-length approximation is used to estimate eddy viscosity and thermal diffusivity;
	\item the evolution of the large-scale magnetic field is described by the induction equation of mean-field theory;
	\item anisotropic diffusion of the magnetic field is considered a substantial factor affecting the overlapping of magnetic cycles and the cycle of dynamo-wave propagation;
	\item $\alpha$-effect is calculated taking into account kinetic and magnetic helicity;
	\item the balance of the large-scale and the turbulent component of magnetic helicity is taken into account.
\end{itemize}

The numerical experiments by Pipin and Kosovichev \cite{Pipin_Kosovichev_2019} based on this model show that there are two necessary conditions for the emergence of an extended 22-year cycle of torsional oscillations. These are, first, the existence of a 22-year cycle of dynamo-wave propagation with overlapping magnetic cycles as observed at the solar surface and, second, the suppression of turbulent heat conduction by the magnetic field with the variation of the meridional circulation in the dynamo cycle as a result. The dynamo wave causes variations in the angular velocity and meridional circulation. 

According to this model, torsional oscillations are due to the influence of the magnetic field on the turbulent angular-momentum transfer and the action of the large-scale Lorentz force. The extended cycle is the effect of the overlapping of successive magnetic cycles combined with the suppression of convective heat transport by the magnetic field and the cyclic variations of meridional circulation in the sunspot-formation zone. 

\section{Conclusion}

The above-described results, obtained by analyzing the data of time--distance helioseismology obtained from the observations of the Solar Dynamics Observatory in 2010--2024, can be summarized as follows.
\begin{itemize}
	\item The characteristic scale of convection increases with depth while its wavelength band narrows. Therefore, this scale is more pronounced in the deep layers than in the upper layers, where flows of a wide range of smaller scales are localized but where the ``depth'' scale is also present. 
		 
	\item The flows observed in the near-surface layers are a superposition of small-scale (``surface'') and large-scale (``depth'') harmonics, the latter covering the entire depth range considered starting from the surface.
	
	\item The observed approach of the spectral peak to the $\ell = m$ line corresponding to sectorial harmonics as we pass to greater depths reflects the tendency of large-scale convective cells to meridional alignment and a banana-like shape.
	
	\item Our analysis for the prolonged time interval shows even more clearly  the previously revealed regularity: the power of convective flows exhibits anticorrelation with the solar-activity level in the near-surface layers and positive correlation in the deeper layers. This indicates the influence of magnetic fields on the intensity of convection: strong fields suppress convective circulation in the near-surface layers and, in contrast, enhance it in deeper layers, possibly due to the quenching of turbulent viscosity.
	
	\item The pattern of the extended, 22-year cycle of zonal flows (``torsional oscillations'') of the Sun, supplemented with the data for the activity-growth phase of Solar Cycle 25, retains its character. 
	          
	\item During the epoch of activity maximum, the meridional flow is superposed with its higher harmonic---a secondary flow converging in the near-surface layers to the latitudinal zone of active-region formation. This feature of the meridional flow, revealed in the helioseismic data of Solar Cycle 24, was also clearly manifested in the Cycle 25.
	
	\item The meridional circulation shows some features of the extended solar cycle, which is consistent with the solar dynamo model \cite{Pipin_Kosovichev_2019}.
\end{itemize}

\bibliographystyle{unsrt}
\bibliography{Getling}
\end{document}